\documentclass[%
 twocolumn,
groupedaddress,
 amsmath,amssymb,
 aps,
]{revtex4-2}

\usepackage{graphicx}% Include figure files
\usepackage{dcolumn}% Align table columns on decimal point
\usepackage{bm}% bold math
\usepackage{xcolor}% for highlighting
\usepackage[flushleft]{threeparttable}
\usepackage{multirow}
\usepackage{longtable}
\usepackage[colorlinks,linkcolor=blue,anchorcolor=blue,citecolor=blue,]{hyperref}
\newcolumntype{.}{D{.}{.}{-1}}
\newcolumntype{d}[1]{D{.}{.}{#1}}
\newcommand*{\wn}{cm$^{-1}$}
\newcommand*{\hsm}{H$_{2}$S}
\newcommand*{\dsm}{D$_{2}$S}
\newcommand*{\Hm}{H$_{2}$}
\newcommand*{\Dm}{D$_{2}$}
\newcommand*{\Hmp}{H$_{2}^{+}$}
\newcommand*{\Dmp}{D$_{2}^{+}$}
\newcommand*{\HDp}{HD$^{+}$}
\newcommand*{\X}{X$^1\Sigma_g^+$}
\newcommand*{\EF}{EF$^1\Sigma_g^+$}
\newcommand*{\E}{E$^1\Sigma_g^+$}
\newcommand*{\F}{F$^1\Sigma_g^+$}

\renewcommand{\eqref}[1]{Eq.~(\ref{#1})}

\begin{document}

\title{Spectroscopic study of the F$^1\Sigma_g^+$ outer well state in H$_2$, HD and D$_2$}

\author{K.-F. Lai}
 \affiliation{Department of Physics and Astronomy, LaserLaB, Vrije Universiteit \\
 De Boelelaan 1081, 1081 HV Amsterdam, The Netherlands}
 %\author{E. J. Salumbides}
 %\affiliation{Department of Physics and Astronomy, LaserLaB, Vrije Universiteit \\
 %De Boelelaan 1081, 1081 HV Amsterdam, The Netherlands}

 \author{M. Beyer}%
 \affiliation{Department of Physics and Astronomy, LaserLaB, Vrije Universiteit \\
 De Boelelaan 1081, 1081 HV Amsterdam, The Netherlands}

  \author{W. Ubachs}%
  \affiliation{Department of Physics and Astronomy, LaserLaB, Vrije Universiteit \\
 De Boelelaan 1081, 1081 HV Amsterdam, The Netherlands}

\date{\today}

\begin{abstract}

Two-photon UV-photolysis of hydrogen sulfide molecules is applied to produce hydrogen molecules in highly excited vibrational levels in the \X\ electronic ground state, up to the dissociation energy and into the quasibound region. Photolysis precursors H$_2$S, HDS and D$_2$S are used to produce vibrationally hot H$_2$, HD and D$_2$.
The wave function density at large internuclear separation is excited via two-photon transitions in the \F\ - \X\ system to probe ro-vibrational levels in the first excited \F\ outer well state of \emph{gerade} symmetry.
Combining with accurate knowledge of the \X($v,J$) levels from advanced ab initio calculations, energies of rovibrational levels in the \F\ state are determined.
For the H$_2$ isotopologue a three-laser scheme is employed yielding level energies at accuracies of $4 \times 10^{-3}$ \wn\ for F($v=0,J$) up to $J=21$ and for some low $J$ values of F($v=1$).
A two-laser scheme was applied to determine level energies in H$_2$ for F($v=0-4$) levels as well as for various F levels in HD and D$_2$, also up to large rotational quantum numbers.
The latter measurements in the two-laser scheme are performed at lower resolution and the accuracy is strongly limited  to 0.5 \wn\ by ac-Stark effects.
For H$_2$ a new quasibound resonance ($v=6$, $J=23$) is detected through the  Q(23) and O(23) transitions in the F0-X6 band.
%Also some quasibound resonances in D$_2$ are tentatively assigned, for the first time in this molecule.
The experimental results on F($v,J$) level energies are compared with previously reported theoretical results from multi-channel quantum-defect calculations as well as with results from newly performed nonadiabatic quantum calculations.

\end{abstract}

\maketitle

\section{Introduction}

The potential energy level structure of the hydrogen molecule, the smallest neutral molecule, is of a peculiar nature.
Strong couplings occur between singly- and double-excited electronic states, with also the ion-pair configuration interacting at large internuclear separation.
The small mass of the hydrogen atom allows for strong adiabatic and nonadiabatic interactions between these electronic manifolds resulting in double-well structures for the potential energy curves of the hydrogen
molecule.

Based on the early spectroscopic studies by Dieke~\cite{Dieke1936,Dieke1949} Davidson concluded that the first excited state of \emph{gerade} symmetry is associated with a double-well potential energy curve~\cite{Davidson1961}.
Later this work was followed by more advanced quantum ab initio calculations of this lowest \EF\ and higher-lying double-well potentials~\cite{Yu1994}.
Precision laser spectroscopic studies focused mainly on rovibrational levels in the \E\ inner well part~\cite{Shiner1993,Zhang2004,Hannemann2006,Salumbides2011,Dickenson2012a,Altmann2018}.

The next higher double-well potential of \emph{gerade} symmetry, denoted as II'$^1\Pi_g$, predicted by Mulliken~\cite{Mulliken1964} and later calculated by ab initio methods~\cite{Dressler1984,Yu1994}, exhibits a shallow outer well some 200 \wn\ below the $n=2$ dissociation limit.
Laser excitation to the large internuclear separation in the outer well was accomplished by multi-step excitation for both H$_2$ and D$_2$~\cite{Reinhold1998} as well as for HD~\cite{Lange2000}.

Higher-lying double-well states of \emph{gerade} symmetry are the GK$^1\Sigma^+_g$ state~\cite{Sprecher2013,Holsch2018} and the H$\bar{\rm H}^1\Sigma^+_g$ state~\cite{Reinhold1997}, which were both investigated via multi-step laser excitation.
Besides this progression of double-well states in the \emph{gerade} manifold of hydrogen there is also a double-well structure in the \emph{ungerade} manifold, the B''$\bar{\rm B}^1\Sigma_u^+$ state, which was excited in a three-laser excitation scheme~\cite{Lange2001,Reinhold1999b}.
Higher-lying double-well states in the \emph{ungerade} manifold were predicted with outer wells at very large internuclear separation~\cite{Detmer1998}.
Attempts to observe rovibrational levels in these higher lying states had limited success~\cite{Koelemeij2003}.

Levels in the \F\ outer well were investigated via emission studies with optical decay from higher lying electronic states exhibiting wave function density at 4-5 $a_0$.
The initial studies by Dieke~\cite{Dieke1936,Dieke1949} were superseded by high-resolution Fourier-transform emission studies both for H$_2$~\cite{Bailly2010} and D$_2$~\cite{Salumbides2014b}, but these were limited to rotational quantum numbers of $J \leq 5$.
Direct two-photon laser excitation from the \X($v=0$) ground state in H$_2$ was pursued to probe levels in the \F\ outer well. Notwithstanding the small Franck-Condon factors this was successful although only levels with $v_F=2,3$ could be detected~\cite{Marinero1983}.

Steadman and Baer found an alternative route for the production of quantum states H$_2$($v$) at large internuclear separation, through the two-photon UV-photolysis of hydrogen sulfide (H$_2$S)~\cite{Steadman1989}.
Detailed studies following this idea were carried out in two- and three-laser schemes disentangling the steps of photodissociation and spectroscopy of the \F\-\X\ system~\cite{Niu2015b,Trivikram2016,Trivikram2019}.
Use of a narrowband ultraviolet laser for the  \F\-\X\ spectroscopy step under Doppler-free conditions led to a test of QED in the highest bound levels X$^1\Sigma_g^+$($v$) in H$_2$, with $v=13$ and $14$~\cite{Lai2021}.
Ultimately, these studies were extended to the region above the X$^1\Sigma_g^+$ dissociation threshold with the detection and precision study of quasibound resonances in H$_2$~\cite{Lai2021b,Lai2021c}.
Quinn et al. explored another photolysis route, choosing formaldehyde (H$_2$CO) as a precursor to produce and study excitation from intermediate vibrational levels \X($v=3-9$) in H$_2$, providing access to the \F\ outer well for levels with $v_F=4$, although in weak transitions~\cite{Quinn2020}.

In the present work the study of F$^1\Sigma_g^+$ - X$^1\Sigma_g^+$ two-photon transitions, involving hydrogen sulfide photolysis as a preparation step, was extended to the HD and D$_2$ isotopologues.
For the first time a comprehensive study of \F\ outer well levels in the mixed HD isotopologue was carried out.
Quadrelli et al.~\cite{Quadrelli1990} had performed an ab initio calculation including non-adiabatic couplings for $J=0-5$ levels, which they had compared with early spectroscopic data for HD~\cite{Dieke1936,Dieke1937}.
Combining the spectroscopic results for F - X two-photon transitions with accurately computed level energies of \X($v,J$)~\cite{Czachorowski2018,Pachucki2022} leads to the determination of F$(v,J)$ level energies, here collected for all three isotopologues.

\section{UV photolysis of hydrogen sulfide}
\label{sec:UV_photolysis}

\begin{figure}
\begin{center}
\includegraphics[width=\linewidth]{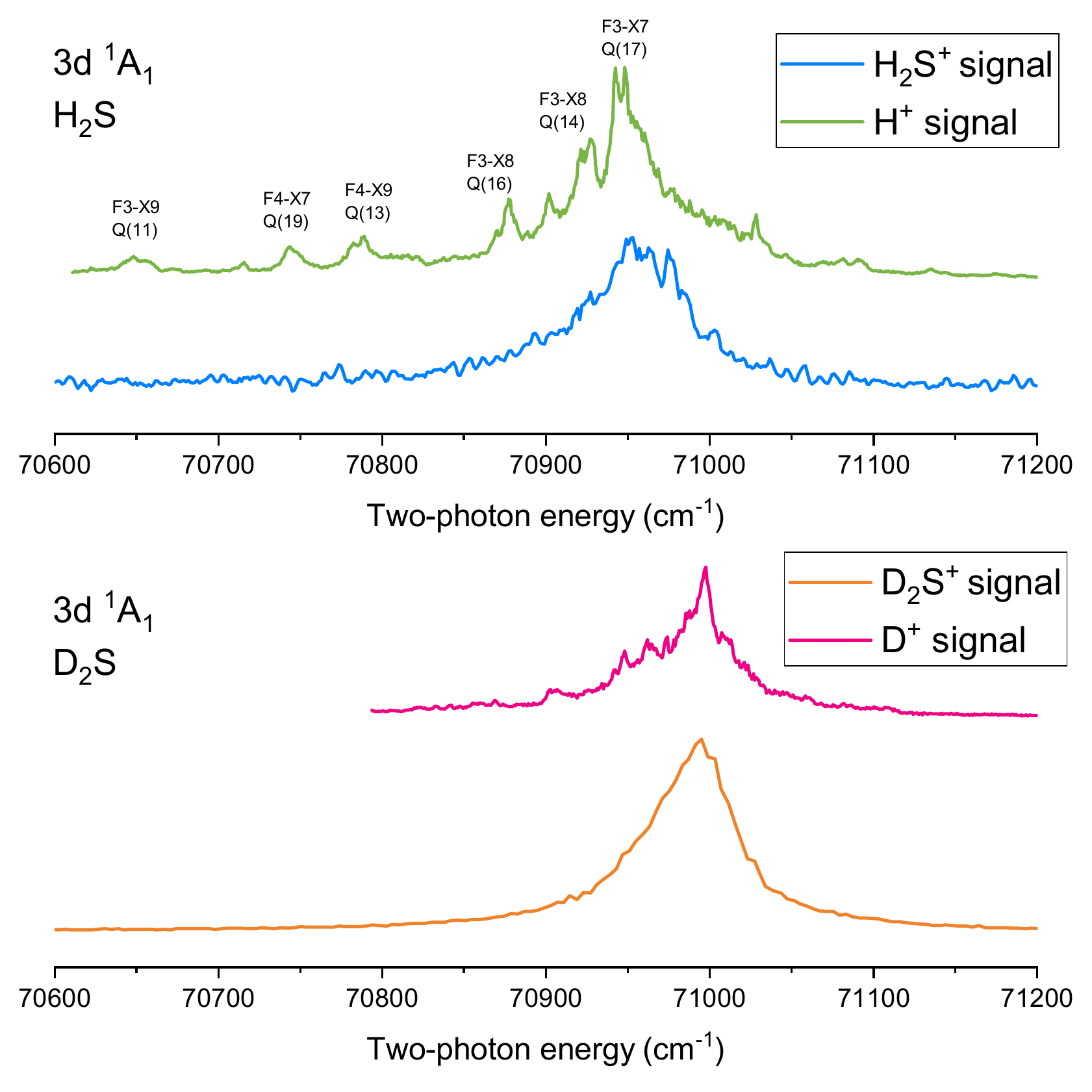}
\caption{\label{H2S-spec}
2+1 REMPI spectra of the \hsm\ and \dsm\ parent molecules indicating the two-photon resonances to $3d$ $^1$A$_1$ electronically excited states used for photoylsis in the present study. Besides parent molecule ionization spectra the H$^+$ and D$^+$ fragment ion traces are also recorded, showing complex structure. Resonances observed in H$^+$ spectra are tentatively assigned with EF-X transitions of~\Hm.
}
\end{center}
\end{figure}

The present study is an extension of previous studies on highly excited ro-vibrational quantum states in \Hm~\cite{Lai2021,Lai2021c}, now also probing highly excited states in HD and \Dm.
Such states are produced via two-photon UV-photolysis of hydrogen sulfide parent molecules.
For two-photon phtolysis at a UV-laser wavelength of 281.8 nm, the $3d\,^1$A$_1$ electronic state is excited in all three isotopologues of the hydrogen sulfide parent molecule.
This follows the assignment of Ashfold et al.~\cite{Ashfold1990}, who had studied the 2+1 REMPI spectra of \hsm\ and \dsm\ in the range of 245-316 nm.
The isotope shift of this electronic band is less than 100~\wn\ for \dsm\ and to be expected smaller for HDS.
In Fig.~\ref{H2S-spec} a 2+1 REMPI spectrum for both H$_2$S and D$_2$S molecules is shown, indicating the frequency positions of the two-photon photolysis resonances in the parent molecules.

In the F-X spectroscopy experiment the wavelength of the photolysis laser was chosen to maximize~\Hmp, \Dmp\ or \HDp\ ion signals from the targeted hydrogen sulfide isotopologues.
In the first stage of the current project the $4p\,^1$A$_1$ resonance in the parent molecule was used (at $\lambda_{\rm UV}=292$ nm) for the photolysis step, providing access to the fragment states H$_2$($v''=11,12$)~\cite{Niu2015b,Trivikram2016,Trivikram2019}.
The main body of the work employed $\lambda_{\rm UV}=281.8$ nm~\cite{Lai2021,Lai2021c}.
Also photolysis pathways were explored through the higher lying $4d\,^1$A$_1$ electronic state with $\lambda_{\rm UV}= 259.9$ nm, as well as the $4f\,^1$A$_1$ state with $\lambda_{\rm UV}= 257.9$ nm.

Following explorative studies by Steadman and Baer~\cite{Steadman1989}, ultraviolet laser photolysis of hydrogen sulfide is employed to produce molecular hydrogen in its ground \X\ electronic state via the pathway:
\begin{equation*}
    \text{XYS} \xrightarrow{2h\nu} \text{S} (^{1}{\rm D}_{2}) + \text{XY},
\end{equation*}
with X, Y being H or D.
The two-photon energy provided for the photolysis, amounting about 70980~\wn\ (for $\lambda_{\rm UV}=281.8$ nm), is sufficient to overcome the dissociation energy of individual hydrogen sulfide isotopologues~\cite{Lai2021,Lai2021c}.
The dissociation energy of \hsm\ in the S($^1$D$_2$) limit has been experimentally determined to be 69935(25)~\wn~\cite{Zhou2020,Zhao2021}.
The 1000 \wn\ of excess energy above dissociation threshold was shown to allow for the formation of quasibound resonances of~\Hm\ at energies up to 500~\wn\ above threshold~\cite{Lai2021b,Lai2021c}.
Ro-vibrational level energies of hydrogen sulfide isotopologues have been calculated~\cite{Tarczay2001,Kozin1994,Miller1990} and found in good agreement with experiment~\cite{Ulenikov1998,Liu2006}.

No experimental values are available for the dissociation energies of \dsm\ and HDS.
These values are estimated following the method of determining the bond dissociation energies of water isotopologues by Boyarkin et al.~\cite{Boyarkin2013}.
The isotope effect on the dissociation energy can be expressed in terms of the zero-point energies ($E_{\rm ZP}$) and the differences of finite-mass corrections $\Delta E_\text{FM}$ of the dissociated atomic species.
The dissociation energy ($D_0$) of \dsm\ and HDS to the S($^1$D$_2$) limit is estimated via:
\begin{equation*}
\begin{split}
    D_0(\text{D$_2$S}) = D_0(\text{H$_2$S}) + E\text{$_{\rm ZP}$(H$_2$S)} - E\text{$_{\rm ZP}$(D$_2$S)} \\
    + 2 \times(\Delta E_\text{FM}(\text H) - \Delta E_\text{FM}(\text D)),\\
    D_0(\text{HDS}) = D_0(\text{H$_2$S}) + E\text{$_{\rm ZP}$(H$_2$S)} - E\text{$_{\rm ZP}$(HDS)} \\
    + (\Delta E_\text{FM}(\text H) - \Delta E_\text{FM}(\text D)),
\end{split}
\end{equation*}
where the finite-mass corrections $\Delta E_\text{FM}$ are about 60 and 30 \wn\ for H and D atoms, respectively.
Table~\ref{tab:diss_erg} lists the $E_{\rm ZP}$ for each isotopologue and the estimated $D_0$ for HDS and~\dsm.
The increased zero-point vibrational energies lead to higher dissocation energies for the deuterated species, and hence to lower excess energies $E_{\rm excess}$, which are evaluated from the difference between the two-photon energy at 70980 \wn\ and $D_0$.
These estimated values set the range of resonance energy for the observable quasibound resonances produced from photolysis.
%The present study focuses on the spectroscopy of the bound states below the dissociation energies of the hydrogen fragment molecules.

\begin{table}
\renewcommand{\arraystretch}{1.3}
\caption{Zero-point- energy ($E_{\rm ZP}$) and dissociation energy ($D_0$) with respect to the S($^1$D$_2$) limit of each hydrogen sulfide isotopologue.
All values in \wn.
\label{tab:diss_erg}}
\begin{threeparttable}
\begin{tabular}{lccc}
Species & $E_{\rm ZP}$ & $D_0$   & \multicolumn{1}{c}{$E_{\rm excess}$}   \\
\hline
\hsm  &   3287    &   69935$^a$    &   1045 \\
HDS   &   2844    &   70408        &   572 \\
\dsm  &   2369    &   70883        &   97 \\
\hline
\end{tabular}
\begin{tablenotes}
\footnotesize
\item $^a$Experimental value from Ref. \cite{Zhou2020}.
\end{tablenotes}
\end{threeparttable}
\end{table}

\section{Experimental procedures}

The experimental configurations for the production of highly excited states in molecular hydrogen and the spectroscopic study of the F$^1\Sigma_g^+$ - X$^1\Sigma_g^+$ transition in a two-laser scheme~\cite{Niu2014} and a three-laser scheme~\cite{Trivikram2016,Trivikram2019,Lai2021,Lai2021c} scheme were presented before.
The measurements on HD and D$_2$ levels were performed in the simplified two-laser scheme with a frequency-doubled grating-based dye laser (LIOPTEC) for the spectroscopy step. A Doppler-free configuration with counter-propagating beams is employed, while the laser bandwidth is about 0.1 \wn\ in the ultraviolet range.
Highly excited rovibrational states of HD and D$_2$ were produced from UV-photolysis and probed via a 2+1 REMPI scheme via the F-X transition, without using a dedicated ionization laser. %tuned to an autoionizing resonance.
The focused UV pulse is configured in a counter-propagating geometry and overlapped spatially with the molecular beam and photolysis laser.
%The molecular hydrogen isotopologues produced are at high vibrational states ($v''$) which have better Franck-Condon overlap with \F\ outer-well states.
This spectroscopy laser scans over the range of 300 - 315 nm.
By this means many transitions were probed exciting \X\ with high $v''$ to \F\ ($v'=0-2$) states, further denoted also as F0, F1 and F2.

Absolute frequency calibration for the spectroscopy step in this two-laser scheme is based on a direct comparison with reference spectra produced via linear absorption spectra of Doppler-broadened resonances in I$_2$~\cite{Gerstenkorn1977} using the visible output of the dye laser at its fundamental wavelength.
The UV-pulse energy of the spectroscopy laser was 1.5 - 2 mJ per pulse.
%to suppress the ac-Stark shift and the unwanted 2+1 REMPI signal from the molecular band of hydrogen sulfide isotopologues.
The pulses of the spectroscopy laser were temporally delayed from pulses from the photolysis laser by 5 ns to avoid ac-Stark contributions by the latter.
While some of the measurements in H$_2$ were performed via this simplfied two-laser scheme, part of the measurements of the F-X transitions in H$_2$ were performed via a more sophisticated three-laser scheme, using a narrowband pulsed-dye-amplification (PDA) system yielding the highest precision~\cite{Trivikram2019,Lai2021,Lai2021c}.

Experiments were performed on pulsed beams of pure \hsm\ and \dsm\ at a stagnation pressure of 2-3 bar.
The HDS sample was produced by pre-mixing \hsm\ and \dsm\ at a ratio of 3.6 : 1.
For an equilibrium constant for the reaction
\begin{equation*}
    \text{H$_2$S} + \text{D$_2$S} \rightleftharpoons \text{2 HDS}
\end{equation*}
of 3.88 at room temperature ~\cite{Wolfsberg1970},
a mixture of~\hsm\ : HDS : \dsm\ of 3.6 : 1.0 : 0.28 is produced.
This mixture is used for the studies of HD spectra.

Signals were recorded by ion detection upon multi-step laser photoionization, involving mass analysis in a time-of-flight (TOF) mass spectrometer, and detection by a multichannel plate with phosphor screen.
For the measurements on~\Hm, the ion optics were triggered at a 80 ns delay from the spectroscopy laser to provide a dc-field-free environment.
Both $m/z = 1$ (H$^+$) or $m/z = 2$ (\Hmp) ions are accelerated into the TOF-tube.
For HD and~\Dm, the background ions from the photolysis process outweigh the signal from F-X transitions.
Therefore a dc-field of 1.3 kV/cm of opposite polarity was applied to remove prompt ions from the photolysis process.
At the accuracy scale of the two-color scheme the induced dc-Stark shift is insignificant.
For  the HD spectra settings at $m/z = 2$ (D$^+$ or~\Hmp) and $m/z = 3$ (\HDp) were employed, where molecular resonances appear in both channels.

\begin{figure}[!b]
\begin{center}
\includegraphics[width=\linewidth]{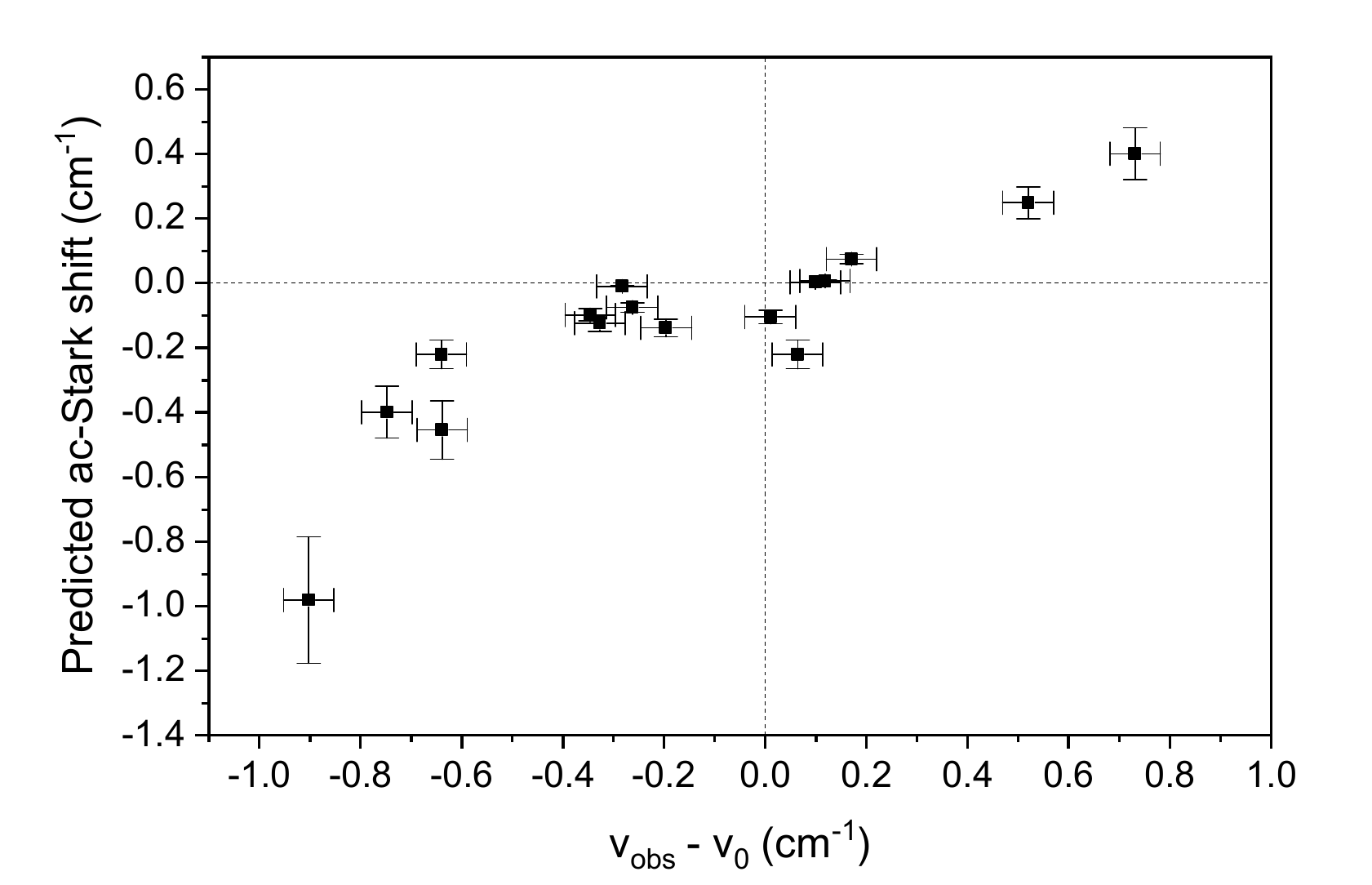}
\caption{\label{H2_acStark}
The predicted ac-Stark shifts for the low-resolution two-color dye-laser measurements, based on measured ac-Stark slope coefficients in the narrowband PDA setup and the power density used in the experiment, plotted against the actually observed ac-Stark shift in the high resolution experiment involving the PDA system. %The data are listed in the Supplementary Material.
}
\end{center}
\end{figure}

The major uncertainty in the calibration of F-X two-photon transition frequencies in the tow-color scheme stems from the ac-Stark effect induced by the spectroscopy laser.
A number of F-X transitions were investigated in the three laser scheme, using the narrowband PDA laser system and performing intensity-dependent measurements with extrapolation to zero intensity.
The latter measurements led to an accuracy of $2 \times 10^{-3}$ \wn\ or 60 MHz for the field-free transition frequencies~\cite{Trivikram2016,Trivikram2019,Lai2021,Lai2021c}.
Examples of such spectra for H$_2$ are presented in the next section.
For this subset of 16 high-resolution measurements on H$_2$ an ac-Stark coefficient was determined from measurements performed at various intensities with extrapolation to zero intensities.
The same set of F-X lines was remeasured in the lower resolution two-laser scheme at a laser energy of 2 mJ/pulse.
The ac-Stark shifts between the measured frequencies $\nu_{\rm obs}$ under these conditions were determined with respect to the zero-intensity values $\nu_0$, while also the expected ac-Stark shifts based on the measured ac-Stark slopes were determined.
Results of a comparison between the observed and predicted values, plotted in Fig.~\ref{H2_acStark}, demonstrates their correspondence.
Taking into account the uncertainties, the correspondence appears to follow a linear behavior; the slope deviating from unity may be ascribed to slightly different focusing conditions and beam geometry in the two- and three-laser schemes.

This result sets a standard for the ac-Stark effect in the F-X transitions, which will be used for estimating uncertainties in the simplified two-color scheme adopted for the HD and D$_2$ spectroscopies.
The root-mean-square of the difference is determined at 0.45~\wn.
Note that in all two-color measurements the laser energy of the UV spectroscopy laser was limited to < 2 mJ/pulse.
Including a smaller uncertainty related to wavelength calibration results in a value 0.5~\wn, defining the experimental uncertainty for the measurements performed in the two-color scheme for all isotopologues.

\section{Spectroscopy of F-X two-photon transitions}

\begin{figure}[!t]
\begin{center}
\includegraphics[width=\linewidth]{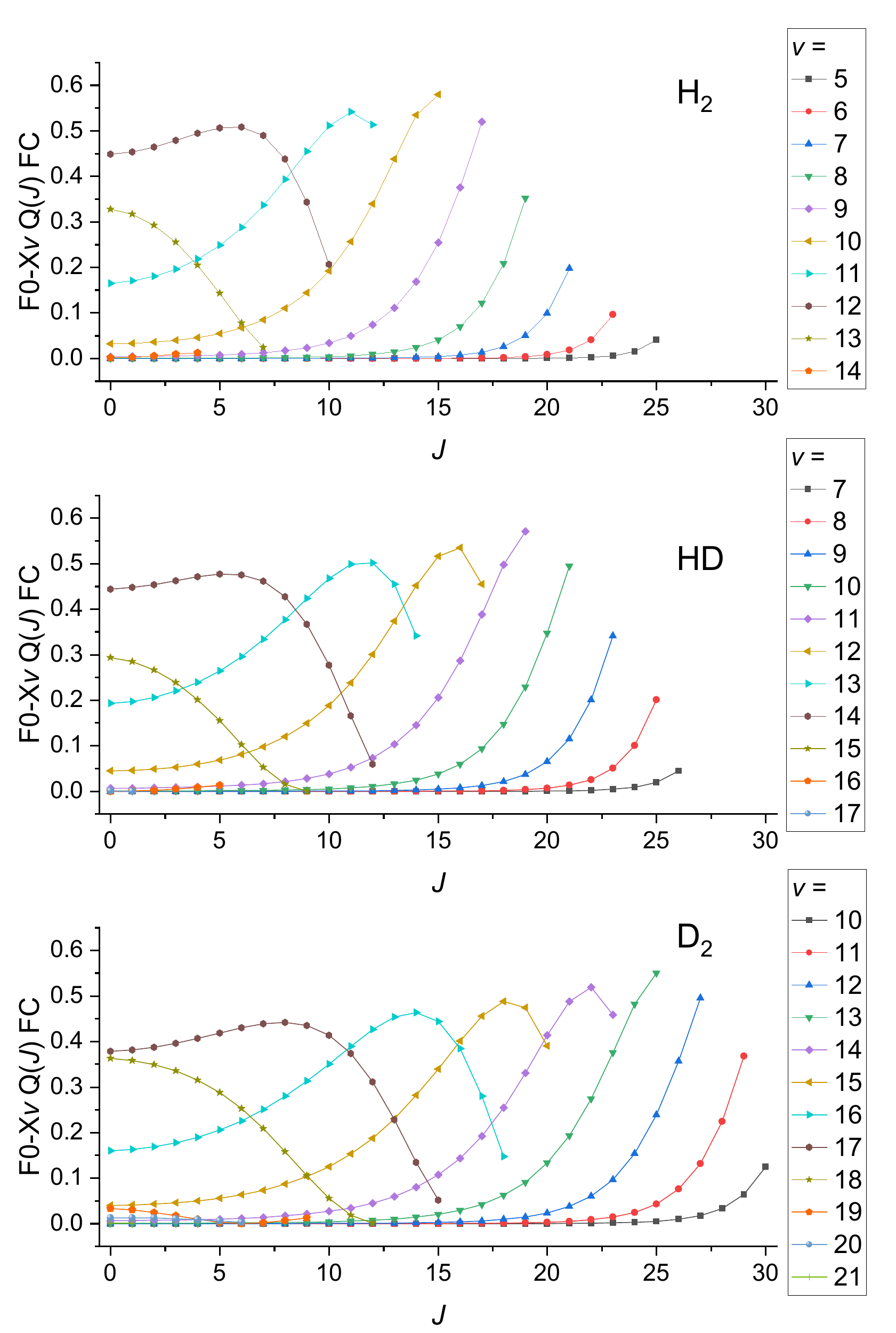}
\caption{\label{FCF-calc}
Calculated Franck-Condon factors for \F\ - \X\ ($0,v''$) Q-branch transitions in~\Hm, HD and \Dm.}
\end{center}
\end{figure}

In this section we present the deduced two-photon transition frequencies in the F$^1\Sigma_g^+$ - X$^1\Sigma_g^+$ system of the hydrogen isotopologues.
The lines in the two-photon spectra were assigned using several ingredients.
Firstly, signal detection was accomplished via the REMPI-TOF method whereby either the molecular fragment ions (H$_2^+$, HD$^+$, D$_2^+$) at masses 2, 3 and 4, or atomic fragments (H$^+$ or D$^+$) at masses 1 and 2, were probed.

Secondly, line intensities were compared to computed Franck-Condon (FC) factors for the various F-X($v',v''$) bands.
For this purpose FC factors were computed for the relevant bands in the F-X system, for Q-branch transitions in the three isotopologues and as a function of rotational quantum number $J$.
Results for the vibrational progression of F-X($0,v''$) bands are presented in Fig.~\ref{FCF-calc}.

\begin{figure}%[!b]
\begin{center}
\includegraphics[width=\linewidth]{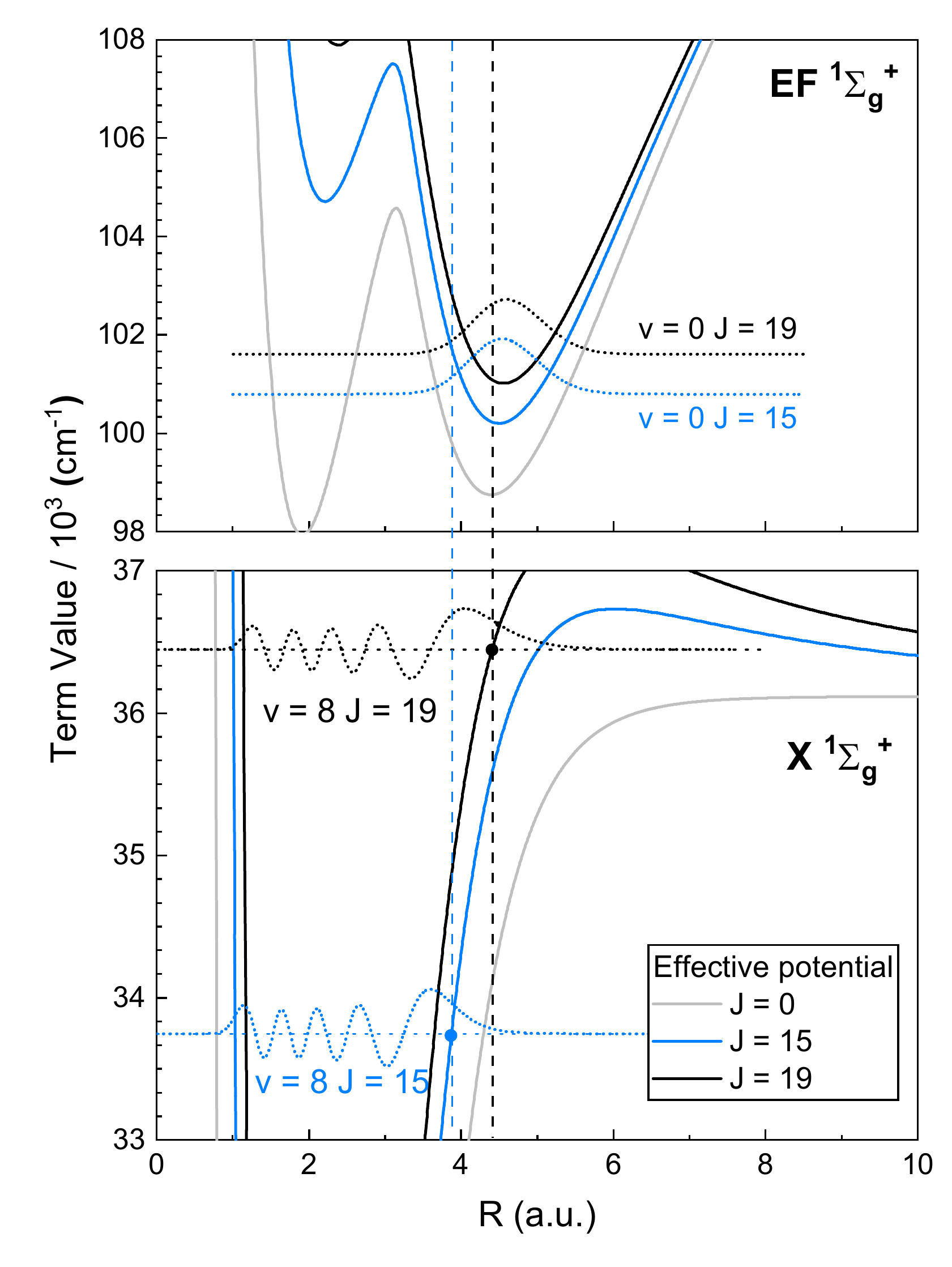}
\caption{\label{wavefunction}
Effective potential energy curves of \X\ and \EF\ state at $J=0, 15, 19$ and radial wavefunction of $v = 8, J = 15$, $v = 8, J = 19$ in \X\ state and $v = 0, J = 15$ and $v = 0, J = 19$ in \F\ state. Dots represent the outer classical turning points for $v = 8, J = 15$ (in blue) and $v = 8, J = 19$ (in black) in \X\ state.
}
\end{center}
\end{figure}

Figure~\ref{wavefunction} shows how the wave function density at the classical outer turning point of the \X\ state potential provides the overlap with the density on the \F\ well.
Vibrational wave functions X($v''=5-14$) thus provide access to the F outer well, but the subtle overlap make the FC-factors strongly $J$-dependent, since the classical outer  turning point extends to larger internuclear distances. As a result, it yields an increasing trend in FC factor with increasing $J$.
For moderately high vibrational levels $v''$, the FC factors increase with the rotational quantum number $J$ for all three isotopologues in the F0-X$v''$ progression. For lower $J$, the wavefunctions of those moderate high $v''$ in X states and F0 state are poorly overlapped and give FC factors less than $10^{-5}$.
The Q-branch transitions with a Franck-Condon factor smaller than $10^{-3}$ were discarded in the analysis, as they may be confused with the weaker S-/O-side-branch lines, which are known to be much weaker than the main Q-branch lines in the \EF-\X\ two-photon transition in molecular hydrogen~\cite{Marinero1983}.

Thirdly, combination differences occurring between lines in the spectra were employed. Combination differences exist between S-/Q-/O lines,
as well as between vibrational splittings in the X-ground and F-excited states.
In the analysis the level energies for the X$^1\Sigma_g^+$($v,J$) ground electronic state as obtained from H2SPECTRE~\cite{SPECTRE2022}, accurate  to within  $10^{-3}$ \wn, are considered to be exact for the purpose of assignment.

As for a fourth ingredient the level energies of \F\ rovibrational states were computed via ab initio methods including non-adiabatic effects.
These calculations use improved BO potential energy curves for excited states of molecular hydrogen that have recently become available~\cite{Silkowski2021}.  The computational methods and results will be presented in section~\ref{NA-calc}.
Alongside multi-channel quantum defect (MQDT) calculations~\cite{Dickenson2012a} were also used to compare with experimental results.

In a few examples, an independent photoionization laser was used for recording autoionization spectra aimed at assigning the rotational quantum numbers of the F-X two-photon transitions
Some examples were presented  previously~\cite{Trivikram2016,Trivikram2019,Lai2021,Lai2021c}.
However, this additional method is limited to the use of the three-color scheme and too time-consuming to apply it to large numbers of spectral lines.
%To assist the assignment, the observed transition was matched to the list of calculated transition frequencies of S-/ Q-/ O-branches with absolute difference less than 5 \wn.

\subsection{H$_2$}

\begin{figure*}[!t]
\begin{center}
\includegraphics[width=1.0\linewidth]{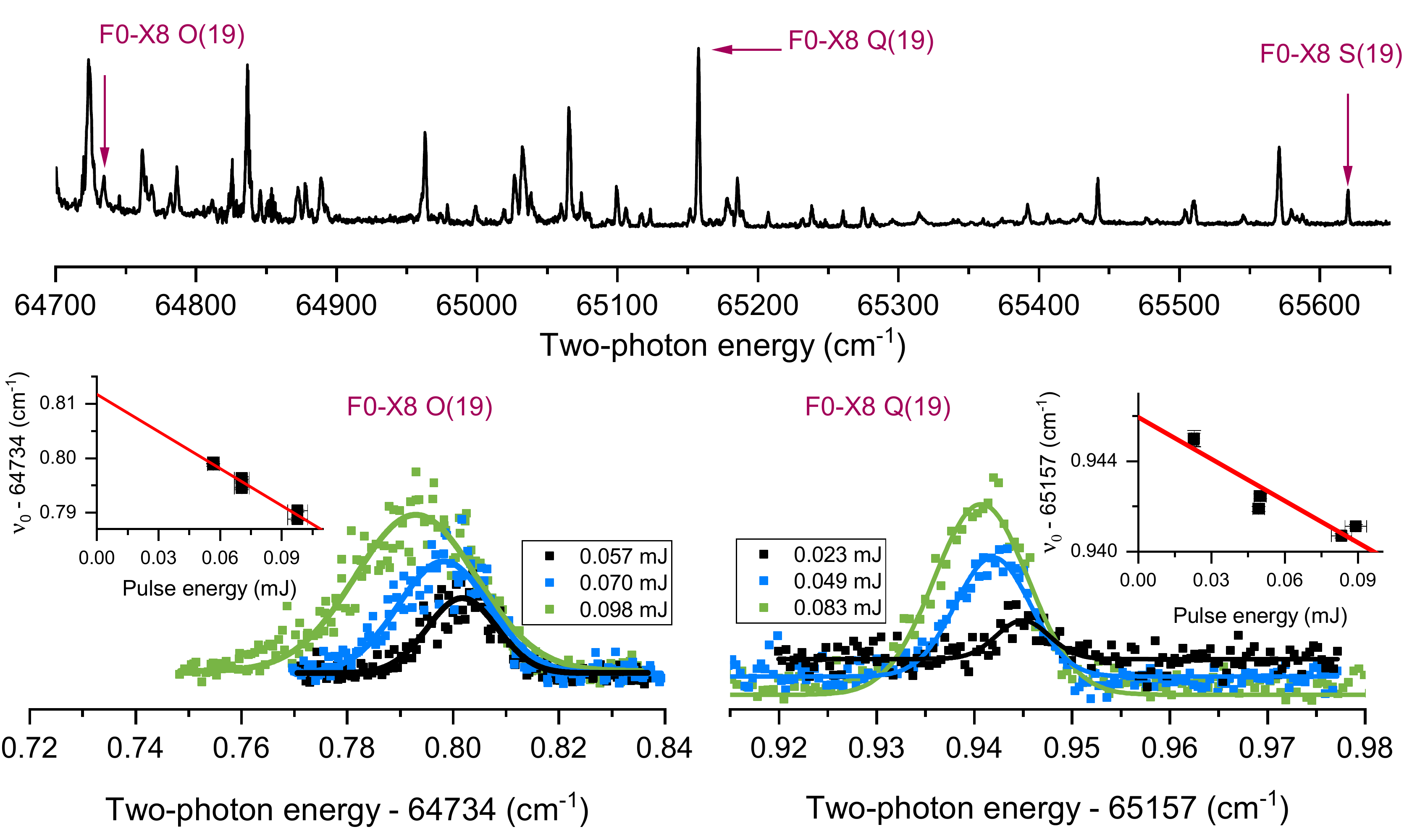}
\caption{\label{H2-spectrum}
Spectra of F-X two-photon transitions in H$_2$ produced via photolysis of \hsm.
Upper panel: Overview low-resolution spectrum recorded via 2+1 REMPI in the two-color scheme. Here all three two-photon allowed transitions from \X\ $v=8, J=19$ to F0 states are observed.
Lower panel: High resolution spectra of two-photon F-X transitions recorded in the three-color scheme, also involving an ac-Stark analysis and extrapolation to zero-intensity of the spectroscopy laser.
}
\end{center}
\end{figure*}

\begin{table}
\caption{Experimentally determined two-photon transition frequencies in the F-X system of H$_2$ with uncertainties given in brackets. The ac-Stark slope coefficients are given in units of \wn/mJ.
}
\begin{tabular}{ccc}
\label{ac-slope}
& & ac-Stark \\
Transition &   $\nu_0$   &  coefficient \\
\hline
F0-X7 Q(21)   & 65441.358\,(1) & 0.12                   \\
F0-X8 Q(19)   & 65157.941\,(2) & -0.06                  \\
F0-X8 S(19)   & 65619.860\,(1) & 0.00                    \\
F0-X8 O(19)   & 64734.810\,(1) & -0.23                 \\
F0-X9 Q(17)   & 64837.297\,(2) & -0.20                \\
F0-X10 Q(15)  & 64493.240\,(1) & 0.20              \\
F0-X8 Q(17)   & 66044.705\,(1) & 0.00                  \\
F0-X9 Q(15)   & 65571.906\,(2) & -0.49                  \\
F0-X9 S(15)   & 65954.451\,(1) & -0.05                      \\
F0-X13 Q(5)   & 63771.984\,(2) & -0.05                    \\
F0-X13 Q(3)   & 63905.931\,(2) & -0.11              \\
F0-X13 Q(1)   & 63993.792\,(2) & -0.07                  \\
F0-X12 Q(4)   & 64781.678\,(1) & -0.11                 \\
F1-X13 Q(0)   & 65207.533\,(5) & -0.00  \\
F1-X13 Q(1)   & 65188.589\,(5) & 0.04  \\
F1-X13 Q(2)   & 65151.813\,(5) & -0.04  \\
\hline
\end{tabular}
\end{table}

Several portions of the 2+1 REMPI spectra of the \F\-\X\ system in \Hm\ were presented in previous publications~\cite{Lai2021,Lai2021c}.
Most of the resonances observed were concentrated between 64\,400 - 65\,200~\wn, covering the range of excitations from X12, X13 and X14 levels to F0 and F1 upper levels. Almost all the transitions observed in \Hmp\ ion spectra were also observed via H$^+$ detection.

In Fig.~\ref{H2-spectrum} an overview spectrum of the F-X two-photon transitions, as measured via the two-laser scheme, is shown for the region 64\,700 - 65\,600 \wn.
Here, three lines are indicated that originate in the same X($v=8,J=19$) ground level.
The assignment yields direct information on the energy separations between rotational levels in the F0 level.
In total 113 lines were observed and assigned in the F($v'$)-X($v''$) system of H$_2$ through measurement of H$^+$ ion spectra, while 44 lines were observed via \Hmp\ ion detection.
In total 118 individual transitions were assigned, in excitation from X($v''=6-14$) levels to F($v''=0-2$).
Some 32 and 10 transitions were observed probing F1 and F2 excited vibrations.
This coverage of quantum states was limited by the scanning range adopted for the spectroscopy laser.

In the lower panels of Fig.~\ref{H2-spectrum} two high-resolution spectra are displayed as recorded in the three-laser scheme exploiting the narrowband PDA system for the spectroscopy step.
A list of results obtained via this high-resolution scheme is presented in Table~\ref{ac-slope}.
For the lines listed the ac-Stark shift was quantitatively analyzed by performing measurements at differing intensities leading to a determination of ac-Stark slope coefficients, as shown in the insets of Fig.~\ref{H2-spectrum}.
The values for the extracted coefficients are given in the Table as experimental parameters in units of \wn/mJ.
In principle these values could be converted to proper power densities in units of ac-Stark shifts in MHz per MW/cm$^2$,
but in view of the large uncertainties in the focal spot size no such conversion will be made and the ac-Stark slope coefficients will only be given in measured units of \wn/mJ.
The regularity of data points in Fig.~\ref{H2-spectrum} indicates that the relative values display a convincing degree of consistency.

The ac-Stark slope coefficients appear to exhibit positive and negative values for the different lines.
The coefficients reflect the entire level structure of the hydrogen molecule since they involve field-induced couplings over all quantum levels in molecular hydrogen~\cite{Trivikram2016}.
AS such the ac-Stark effect may be considered a high-order moment of the quantum level structure of a molecule.
No quantitative theory for these ac-Stark shifts and slope coefficients has been developed.
Development of such a theory, tested against experimental parameters would provide information on dynamical polarizabilities of the hydrogen molecule~\cite{Trivikram2016}.
In the literature values have been presented for ac-Stark slope coefficients over several decades, in many cases focusing on the EF-X two-photon transition in H$_2$~\cite{Srinivasan1983,Vrakking1993,Yiannopoulou2006,Dickenson2013,Niu2014,Trivikram2016}.
All resulting values are of order few MHz per MW/cm$^2$, but they appear randomly distributed.
The listing of the present large data set on slope coefficients may be useful for benchmarking future theories on the phenomenon.

\begin{figure}[!t]
\begin{center}
\includegraphics[width=\linewidth]{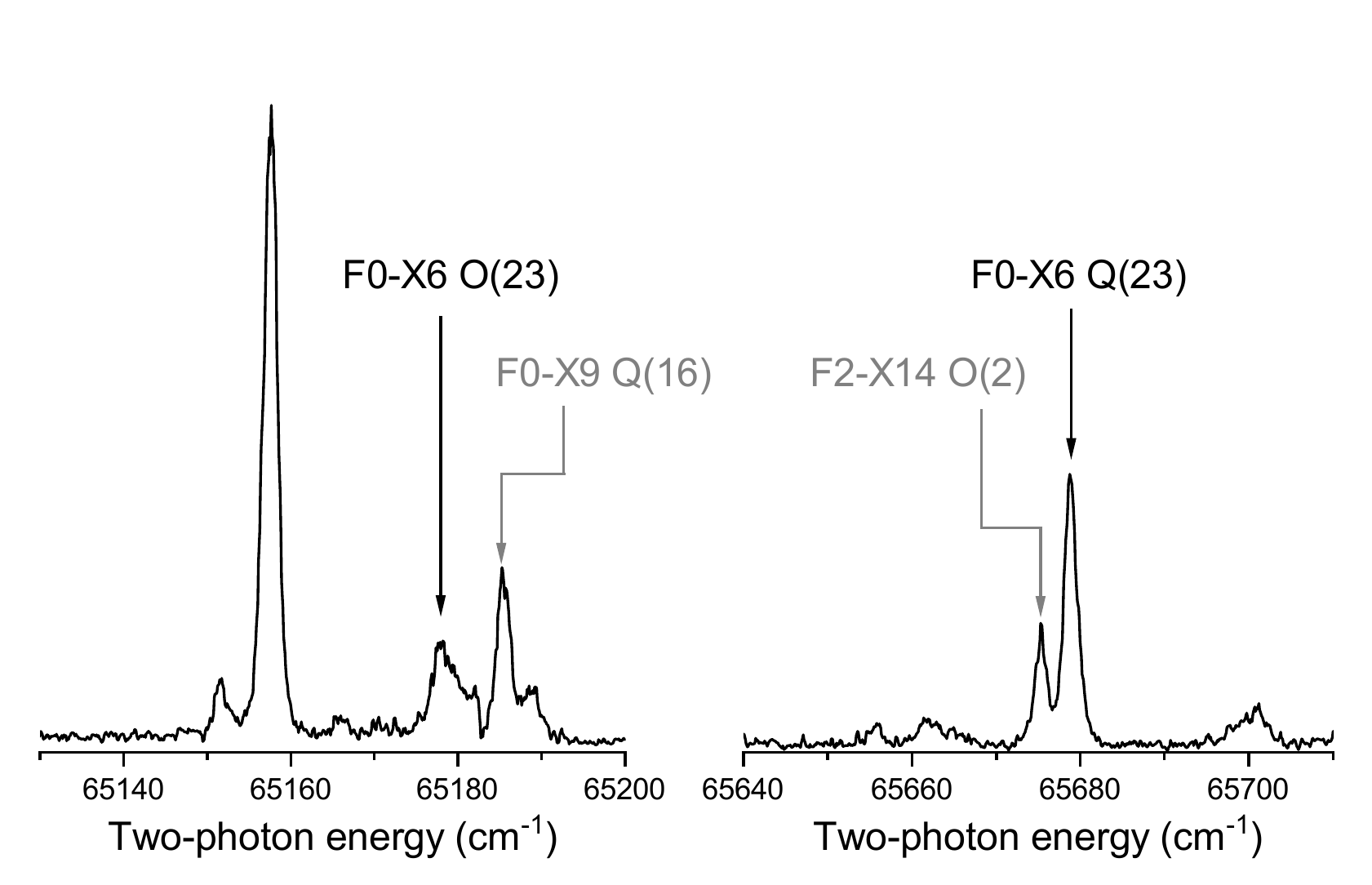}
\caption{\label{X6J23}
Observation of the quasibound resonance X($v$, $J$)=(6,23) excited via the F0-X6 Q(23) and the weak F0-X6 O(23) two-photon transitions, using H$^+$ ion detection.
}
\end{center}
\end{figure}

In previous studies~\cite{Lai2021b,Lai2021c}, five quasibound resonances on the X-ground state potential have been observed through F0-X transitions: ($v,J$) = (11,13), (10,15), (9,17), (8,19), (7,21).
In the present two-color spectra, four of these quasibound resonances, except for (11,13), are again observed in both F0-X and F1-X Q-branch transitions.
The absence of the (11,13) quasibound resonance in the two-color spectra may be due to its short 1.6 ns lifetime.
In the present study an additional quasibound resonance (6,23) is found through the F0-X6 Q(23) transition at 65678.80 \wn\ as shown in Fig. \ref{X6J23}.
The calculated excess energy of the (6,23) resonance is 767.79 \wn, while its lifetime is about 30 $\mu$s~\cite{Selg2012}.
As discussed in Section \ref{sec:UV_photolysis}, (6,23) is energetically favourable to be produced upon photolysis of \hsm\ at 281.8 nm.
Through the combination difference with the F0-X7 S(21) transition, the ground state interval between (7,21) and (6,23) is 261.15~\wn, while the calculated value is 261.86~\wn.
%Further proof of the assignment should be provided from the autoionization spectra. However, these transitions have never been measured successfully with high resolution measurements.

\begin{figure*}
\begin{center}
\includegraphics[width=\linewidth]{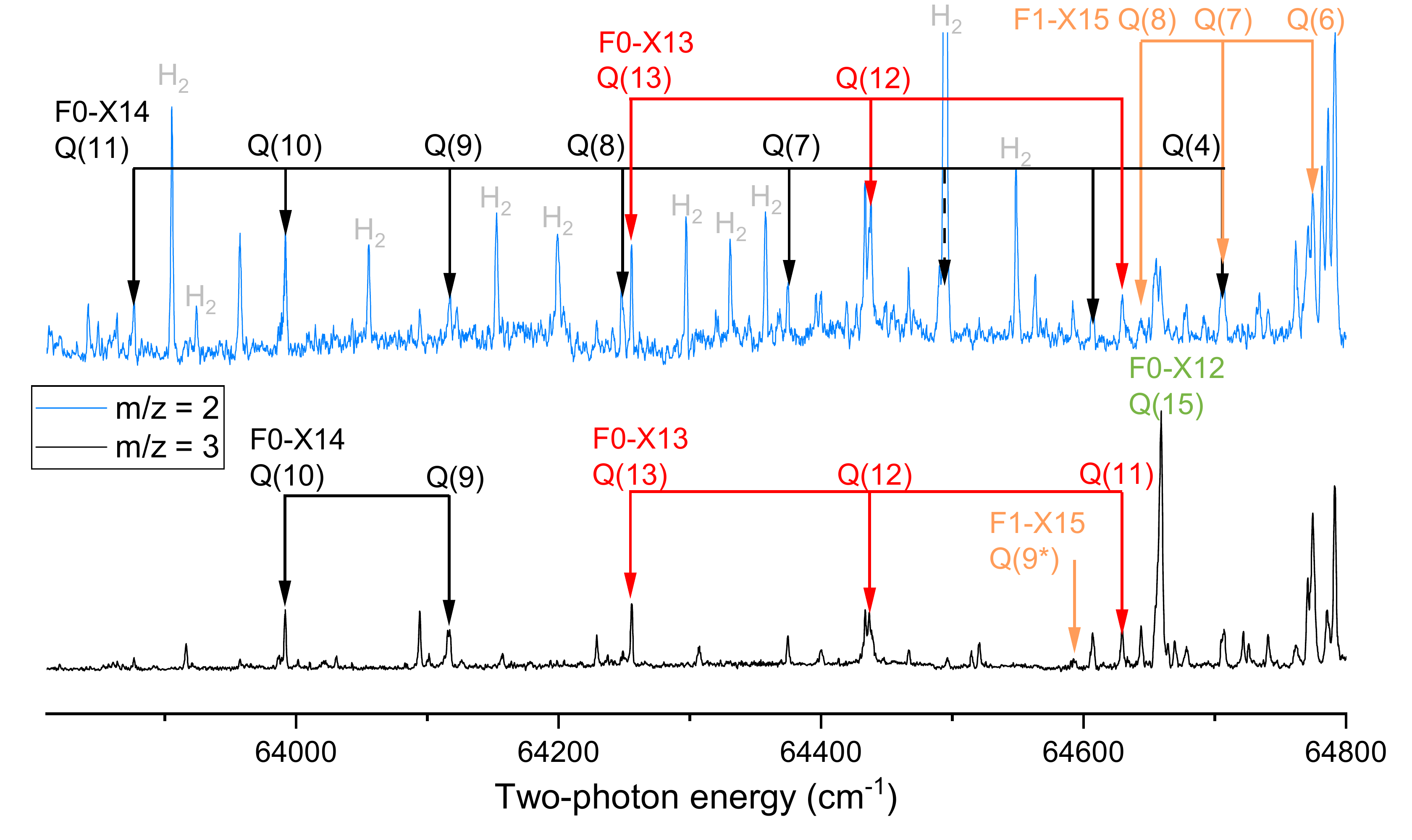}
\caption{\label{HD_overview}
Overview of two-photon spectra in the F-X system in for the HD photofragment as recorded in the two-color laser scheme with the H$_2$S/HDS/D$_2$S premixture before photolysis.
The upper panel is the spectrum for $m/z=2$ (D$^+$ and H$_2^+$ detection; the lower panel for $m/z=3$ detection (HD$^+$).
}
\end{center}
\end{figure*}

The large set of data on the low-resolution \F\-\X\ two-photon transition frequencies in H$_2$ are listed in Table~\ref{H2list} in the Appendix. For these data the major uncertainty derives from the estimated ac-Stark effect.

%The calculated Franck-Condon factors of E($v'=0,1$) - X($v''>5$) transitions are generally smaller than $10^{-4}$ as most of the bound states of the \EF\ electronic state are localized in either the inner or outer well potential.
%The tunneling effect between the E and F states could significantly mix the character of E/F states and via this mechanism allow for E-X transitions originating in high vibrational levels X($v''$) for specific $J$ values.
%Such tunneling mixing, involving E1($J=2,3$) levels, was observed previously~\cite{Marinero1982}. A similar mixing is predicted between E0 and F1 states at $J=7$, but no related transition has been observed in the scanning range of the present experiment.

\subsection{HD}

\begin{figure*}
\begin{center}
\includegraphics[width=\linewidth]{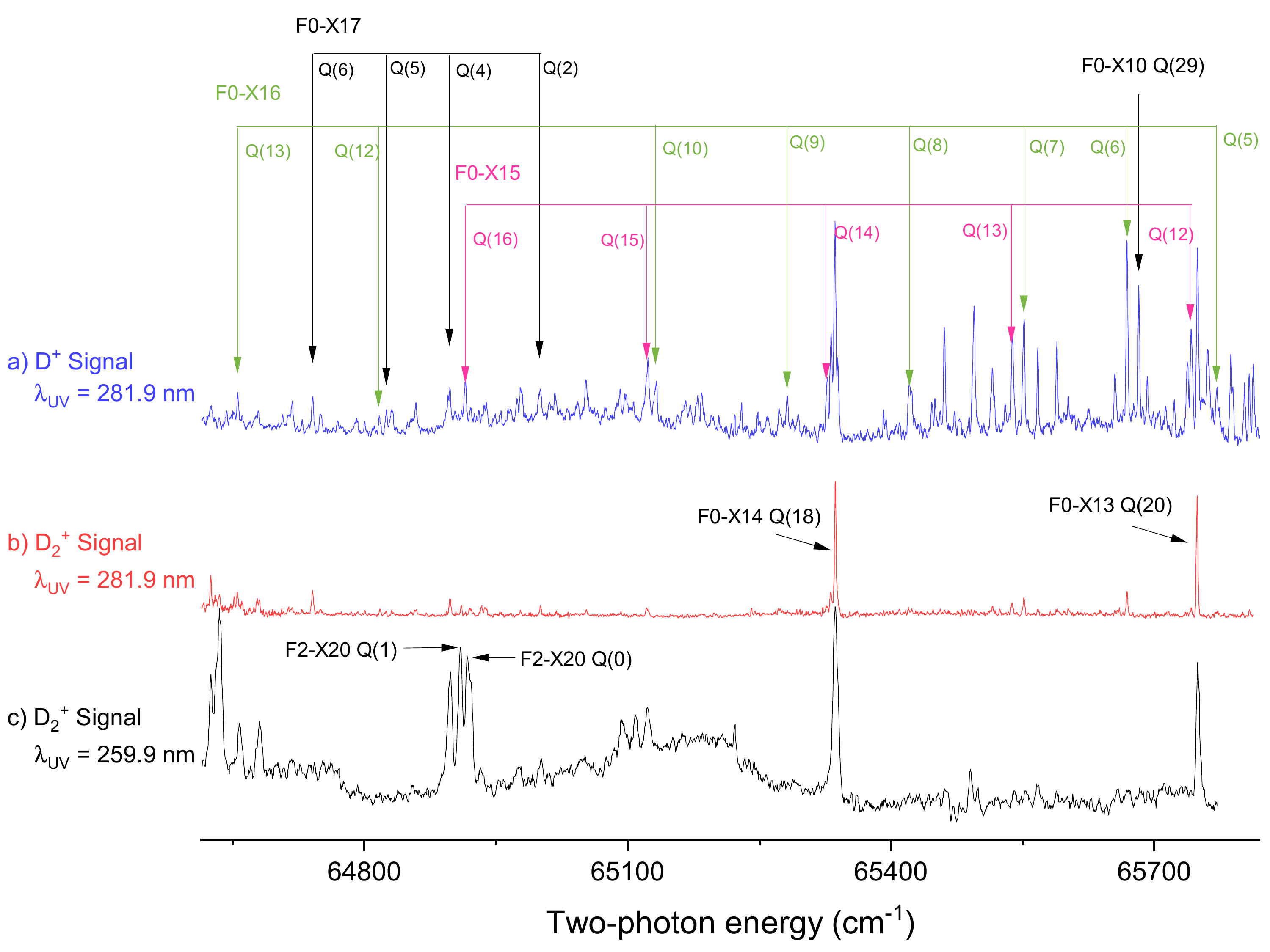}
\caption{\label{D2_overview}
Overview spectra of D$_2$ resonances in the F-X system upon photolysis of \dsm\ for the various choices of photolysis wavelengths and fragment ion detection as indicated.
}
\end{center}
\end{figure*}

Spectra of HD were recorded using a premixture of H$_2$S/HDS/D$_2$S gases for the molecular beam.
The (2+1) REMPI spectra of the \F\ - \X\ transitions were recorded through H$^+$, D$^+$ and HD$^+$ signal detection.
The H$^+$ detected spectra were overwhelmed by the (2+1) REMPI signal from \Hm\ produced from the gas mixture with over 70\% of ~\hsm.
The $m/z = 2$ signals are produced by \Hmp\ ions from~\Hm\ and D$^+$ from HD or~\Dm.
Fortunately, these \Hmp\ signals from \Hm\ are sparse in the current scanning range, while the D$^+$ signals from \Dm\ are of low intensity as only 5\% of \dsm\ is contained in the precursor mixture.
For the analysis of F-X resonances in HD (2+1) REMPI spectra are recorded in separate scans with the ion-mass analyzed set at $m/z = 2$ (probing D$^+$ and H$_2^+$ ions) and at $m/z = 3$ (probing HD$^+$).
In contrast to the case of \hsm\, large non-resonant background ion signals were detected from the photolysis process.
In order to mitigate this and to increase the signal-to-background for the HD spectra, a 1.3 kV/cm dc-field was applied to remove prompt ions.

Fig.~\ref{HD_overview} shows the HD spectrum in the scan region 63\,800 - 64\,800 \wn\ for both  $m/z = 2$ and  $m/z = 3$ ion traces.
Many transitions probed at $m/z = 2$ can be assigned to H$_2$ as indicated in the top panel, while a good fraction of the lines is assigned to HD resonances. For HD mostly lines in the F0-X13, F0-X14 and F1-X15 bands are identified.
In total 97 lines and 108 lines were observed and assigned in the HD F-X transitions through D$^+$ and HD$^+$ ion spectra respectively.
In total 129 individual transitions were labeled with X($v''=7-17$) levels to F($v'=0-3$) levels.
Similar as in the case of \Hm, the majority of transitions were assigned as F0-X, amounting to 91 lines observed.
Only 10 transitions to F($v'=2,3$) levels were assigned by the limited spectroscopic laser scanning range and the relatively small Franck-Condon factors, compared to F0-X and F1-X systems.
A complete list of transition frequencies in HD is presented in Table~\ref{HDlist} in the Appendix.

\begin{table*}
\caption{\label{tab:Fstate-H2}
Derived experimental level energies of F0 and F1 states in H$_2$ measured in the high resolution setup using the PDA laser system. A comparison is made with data from FT-emission spectra~\cite{Bailly2010}, results from MQDT calculations~\cite{Dickenson2012a} as well as from the present non-adiabatic calculations. All values in \wn.
}
\begin{tabular}{cc.....}
 & $J$ & \multicolumn{1}{c}{Exp.} &  \multicolumn{1}{c}{FT~\cite{Bailly2010}}  & \multicolumn{1}{c}{MQDT~\cite{Dickenson2012a}}  & \multicolumn{1}{c}{NA}    \\
\hline
\multirow{15}{*}{F0} &  0 & 99363.805(5) &  99363.8875(4) & 99365.22 & 99363.50\\
  &  1  & 99376.044(4) &  99376.0474(4) & 99377.46  & 99375.75 \\
  &  2  & 99400.504(4) &  99400.512(1)  & 99401.92 & 99400.24 \\
  &  3  & 99437.160(4) &  99437.1665(5) & 99438.57 & 99436.94 \\
  &  4  & 99485.964(4) &  99485.971(3)  & 99487.37 & 99485.79 \\
  &  5  & 99546.859(5) &  99546.868(1)  & 99548.26 & 99546.76\\
  &  6  & 99619.784(6) &       -         & 99621.16 & 99619.75 \\
  &  7  & 99704.622(3) &       -         & 99706.04 & 99704.68 \\
  &  9  & 99909.63(2)  &       -         & 99911.06 &  99909.93\\
  &  11 & 100160.993(5) &     -          & 100162.41 & 100161.54\\
  &  13 & 100457.50(2)  &     -          & 100458.95 & 100458.33\\
  &  15 & 100797.766(4) &     -          & 100799.23 & 100798.94\\
  &  17 & 101180.310(4) &     -         & 101181.92 & 101181.84\\
  &  19 & 101603.441(5) &    -          & 101604.93 & 101605.35\\
  &  21 & 102065.360(4) &    -          &    -      & 102067.66 \\
\hline
\multirow{4}{*}{F1} &  0 & 100558.853(5) &  100558.851(1)  & 100563.13 & 100558.95 \\
  &  1 & 100570.841(5) & 100570.843(3)  & 100575.13 & 100570.58 \\
  &  2 & 100594.800(5) &  100594.8070(6) & 100599.11 & 100594.57 \\
  &  4 & 100678.50(2)  &  100678.510(1)  & 100682.87 & 100678.37 \\
\hline
\end{tabular}
\end{table*}

\begin{figure}[!t]
\begin{center}
\includegraphics[width=\linewidth]{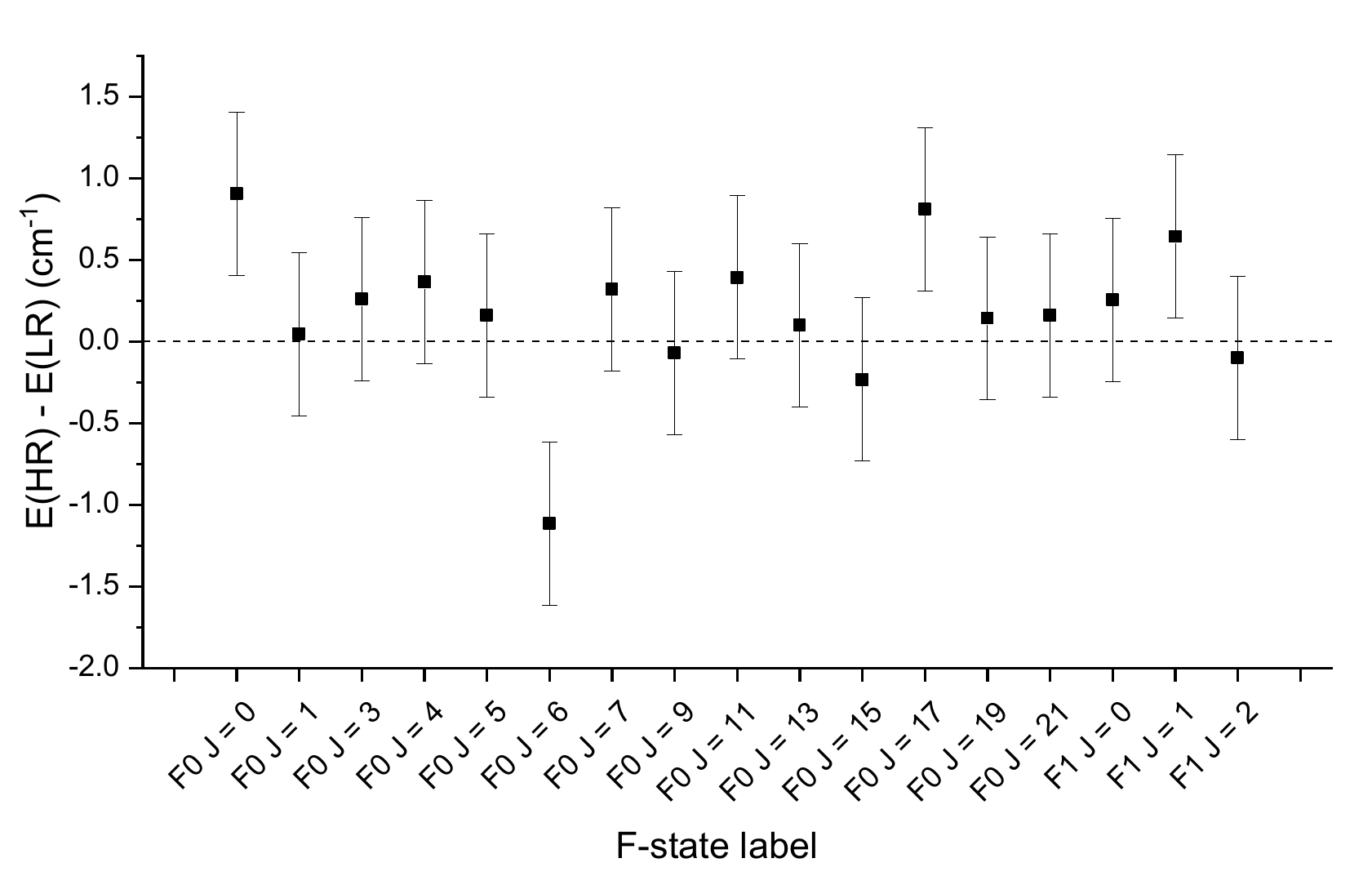}
\caption{\label{H2_HRLR_comparison}
Difference between derived level energies for levels in the \F\ state in \Hm\ measured via the high resolution setup, resulting in $E$(HR), and via the low resolution two-color setup, resulting in values $E$(LR).}
\end{center}
\end{figure}

\subsection{D$_2$}\label{sec:D2_F-X}

The quality of \Dm\ spectra is inferior to those of the other isotopologues, even though pure \dsm\ was used as precursor in the photolysis.
The photolysis of \dsm\ was found to generate large amounts of prompt ions, which were partly removed by applying an opposite polarity dc-field, but still the remaining background lowers the sensitivity of the (2+1) REMPI signals for~\Dm.
The scanning range of the spectroscopy laser also covers several absorption bands of the ~\dsm\ parent molecule, for example the 4$p$ $^1$A$_2$ band at 315 nm and vibronic bands at 302 and 310 nm.
Hence the spectroscopy lasers also cause photolysis and the production of various fragment ions.
Reducing the pulse energy of the spectroscopy laser was not an option since the 2+1 REMPI signals of \Dm\ would lower the signal-to-noise by too much.

In Fig.~\ref{D2_overview} spectra of F-X two-photon transitions in the D$_2$ molecule are presented. Three different recordings were performed over the spectral range 64\,650 - 65\,750 \wn.
With the dissociation laser set at $\lambda_{\rm UV}=281.8$ nm many resonances were observed for detection in the D$^+$ signal channel ($m/z=2$).
This spectrum shows the progressions F0-X($v'$) for $v'=15-17$.
For detection in the D$_2^+$ channel ($m/z=4$) only a few marked resonances were observed in this energy range.

In total 93 lines were observed via D$^+$ ion detection and 39 lines via D$_2^+$ ion detection.
In total 112 individual transitions were assigned to F-X two-photon transitions combining X($v''=10-20$) ground levels and F($v'=0-4$) excited levels.
The spectra in the \Dm\ measurement appear denser than those for  \Hm\ and HD due to its larger moment of inertia.
This density limits the assignment of transitions in regions where unresolved bands are observed at poor signal-to-noise ratio.
The fact that the lines observed via D$_2^+$ detection are erratically distributed may point to a decisive influence of the ionization step, whereby signal is only enhanced for some random lines, where the ionization step matches a transition from the F-state to an autoionizing resonance in the two-color scheme.

In a third spectrum the photolysis laser was moved to $\lambda_{\rm UV}=259.9$ nm, while the ion detection was kept at the D$_2^+$ channel.
Some strong transitions for photolysis at $\lambda_{\rm UV}=281.8$ nm are reproduced, while some additional resonances probing lines in the F2-X20 band appear in the spectrum.

From Table \ref{tab:diss_erg}, the estimated $E_{\text{excess}}$ for photolysis of \dsm\ at 281.8 nm is about 97~\wn, which limits the number of quasibound resonances plausible to form. In the present studies, four quasibound resonances, $(v,J) = $(13,24), (15,20), (16,18) and (17,15), were observed with resonance energies at about 190, 164, 179 and 31~\wn, respectively. (15,20) and (17,15) resonances were observed with Q-branch transitions via F0-X15 Q(20) at 64181.15~\wn, F1-X17 Q(15) at 64624.84 \wn\ and F2-X17 Q(15) at 66507.91~\wn. However, (13,24) and (16,18) were only observed through the S-/ O-branch transitions. Among the four quasibound resonances, only (17,15) have resonance energy below the $E_{\text{excess}}$ of \dsm\ photolysis. The estimated $E_{\text{excess}}$ for \dsm\ photolysis has 25 \wn\ uncertainty limited by the experimentally determinated dissociation energy of~\hsm. From the measurement of \dsm\ photolysis at 259.9 nm with $E_{\text{excess}} > 6000$~\wn, F1-X17 Q(15) transition endured in the spectra. However, further proof is needed for the correct assignment of these transitions.

The observed two-photon transition frequencies in the F-X system of D$_2$ are presented in Table~\ref{D2list} in the Appendix.

\section{F-level energies}

The term values for the \F\ outer well rovibrational levels, relative to the \X($v=0, J=0$) ground level, can be derived by adding the presently measured two-photon transition frequencies to the calculated term values of ro-vibrational \X\ levels~\cite{Czachorowski2018,Pachucki2022}.
In this procedure a statistical and weighted average is taken over all calibrated two-photon lines probing the same upper level in the \F\ state.
A similar approach was followed for the determination of level energies for high rotational states in the E0 - E3 inner well  states with $J$ up to 19~\cite{Dickenson2012a}.
The accuracy of the X($v,J$) ground state levels is about $10^{-3}$ \wn, for some levels at low $v$ and $J$ even as accurate as $10^{-5}$ \wn~\cite{Holsch2019,Puchalski2019b}.
The accuracy for the highest X($v$) levels was experimentally tested via measurement of combination differences for $v''=13,14$~\cite{Lai2021}.

\begin{table}
\caption{\label{tab:Fstate-H2-lowres}
Derived experimental level energies of F0, F1 and F2 states in \Hm\ measured via the lower accuracy two-laser scheme.
The accuracy of the level energies of 0.5 \wn\ is limited mainly by the (estimated) ac-Stark effect. A comparison is made with results from the present non-adiabatic calculations and with results from MQDT calculations~\cite{Dickenson2012a}. All values in \wn.}
\begin{tabular}{cc...}
State & $J$ & \multicolumn{1}{c}{Exp.}& \multicolumn{1}{c}{NA. Calc.}  &  \multicolumn{1}{c}{MQDT~\cite{Dickenson2012a}}\\
\hline
\multirow{8}{*}{F0} & 8  & 99801.2(5)  & 99801.44 & 99802.69\\
 & 10 & 100029.4(5) & 100030.01 & 100031.02\\
 & 12 & 100303.9(5) & 100304.37 & 100305.16\\
 & 14 & 100622.4(5) & 100623.25 & 100623.71\\
 & 16 & 100984.0(5) & 100985.21 & 100985.51\\
 & 18 & 101386.2(5) & 101388.63 & 101388.55\\
 & 20 & 101829.5(5) & 101831.78 & -\\
 & 23 & 102564.3(5) & 102566.84 & -\\
\hline
\multirow{15}{*}{F1} &  3  & 100629.8(5) & 100630.52 & 100635.04 \\
&  5  & 100737.5(5) & 100738.06 & 100742.53 \\
&  6\footnote{Tentative assignment}  & 100811.6(5) & 100809.51 & 100813.97\\
&  7  & 100892.2(5) & 100892.66 & 100897.02\\
&  8  & 100986.8(5) & 100987.29 & 100991.72\\
&  9  & 101093.5(5) & 101093.39 & 101097.75\\
&  10 & 101210.7(5) & 101210.78 & 101215.08\\
&  12 & 101477.7(5) & 101478.82 & 101483.13\\
&  13 & 101628.4(5) & 101629.14 & 101633.31\\
&  14 & 101789.0(5) & 101790.09 & 101794.16\\
&  15 & 101960.1(5) & 101961.46 & 101965.41\\
&  16 & 102141.8(5) & 102143.06 & 102147.13\\
&  17 & 102333.1(5) & 102334.68 & 102338.46\\
&  19 & 102745.1(5) & 102747.07 & 102750.68\\
&  21 & 103194.4(5) & 103196.79 & -\\
\hline
\multirow{9}{*}{F2}&  0  & 101698.5(5) & 101698.74 & 101703.92 \\
&  1  & 101710.9(5) & 101710.65 & 101715.79 \\
&  2  & 101734.6(5) & 101734.87 & 101739.80 \\
&  3  & 101768.4(5) & 101768.45 & 101773.47\\
&  5  & 101874.3(5) & 101874.84 & 101879.68\\
&  6  & 101943.2(5) & 101945.09 & 101949.82 \\
&  7  & 102026.8(5) & 102026.74 & 102031.26\\
&  9  & 102224.7(5) & 102223.76 & 102227.83\\
&  11 & 102465.0(5) & 102464.72 & 102468.22\\
\hline
\end{tabular}
\end{table}

Table~\ref{tab:Fstate-H2} lists the thus derived \F\ state term values in ~\Hm\ as obtained from the accurate three-laser experiments employing the narrow PDA system for the spectroscopy step, while also the ac-Stark effect was addressed quantitatively.
Such high-accuracy values are determined for F0 and rotational quantum numbers $J=0-21$, restricted to low $J$ and for higher-$J$ levels to the odd $J$ levels in ortho-\Hm.
In addition four term values in F1 were determined for low-$J$ values.
Table \ref{tab:Fstate-H2} also includes a comparison with experimentally determined energies of F0 and F1 levels form Fourier-transform emission studies~\cite{Bailly2010} as well as with a multi-channel quantum-defect (MQDT) calculation for the same levels~\cite{Dickenson2013} up to $J=19$.
%The comparison between the present data with MQDT calculation shows an averaged $-1.4$ \wn\ offset throughout the entire $J$-manifold for F0. For the F1 levels the difference is  $-5$ \wn.
The differences with the results from the FT-emission data~\cite{Bailly2010} are all less than $0.01$ \wn\ and at the level of the combined accuracies, with the exception of the F0($J=0$) level where the difference is much larger amounting to $0.08$ \wn.

\begin{table}
\caption{\label{tab:Fstate-HD-F0}
Derived experimental level energies of F0 states in HD measured in this work. A comparison is made with the present non-adiabatic calculations.
}
\begin{tabular}{cc..}
State & $J$& \multicolumn{1}{c}{Exp.}& \multicolumn{1}{c}{NA. Calc.}      \\
\hline
\multirow{26}{*}{F0} &  0  & 99577.5(5)  & 99576.54  \\
 &  1  & 99586.4(5)  & 99585.74  \\
 &  2  & 99604.3(5)  & 99604.15  \\
 &  3  & 99632.1(5)  & 99631.73  \\
 &  4  & 99668.8(5)  & 99668.47  \\
 &  5  & 99714.0(5)  & 99714.35  \\
 &  6  & 99769.7(5)  & 99769.28  \\
 &  7  & 99833.2(5)  & 99833.26  \\
 &  8  & 99907.4(5)  & 99906.21  \\
 &  9  & 99987.9(5)  & 99988.06  \\
 &  10 & 100078.5(5) & 100078.75 \\
 &  11 & 100178.1(5) & 100178.20 \\
 &  12 & 100286.4(5) & 100286.31 \\
 &  13 & 100402.9(5) & 100402.99 \\
 &  14 & 100527.5(5) & 100528.14 \\
 &  15 & 100661.1(5) & 100661.66 \\
 &  16 & 100803.0(5) & 100803.42 \\
 &  17 & 100953.5(5) & 100953.31 \\
 &  18 & 101110.5(5) & 101111.21 \\
 &  19 & 101276.0(5) & 101276.98 \\
 &  20 & 101449.4(5) & 101450.48 \\
 &  21 & 101630.6(5) & 101631.58 \\
 &  22 & 101818.5(5) & 101820.14 \\
 &  23 & 102014.9(5) & 102015.99 \\
 &  24 & 102217.4(5) & 102218.98 \\
 &  26 & 102643.1(5) & 102645.78 \\
\hline
\end{tabular}
\end{table}

In addition to the high-accuracy data extracted from the three-color PDA-based measurements, level energies of some missing rovibrational levels in F0 and F1 states, as well as for F2 were derived from the lower resolution two-color laser measurements.
Those results are presented in Table~\ref{tab:Fstate-H2-lowres}.
The energies for the F-levels are obtained by averaging over results of measurements of multiple lines in the F-X system.
For a subset of 16 levels results for F-level energies as obtained from the high-resolution three-color scheme are compared to results as obtained form the low-resolution two-color laser excitation scheme.
The deviations, as plotted in Fig.~\ref{H2_HRLR_comparison}, show that those fall again statistically within the  $1\sigma$ uncertainty of 0.5 \wn.
This proves that the methods of the two-color laser scheme yield F-level energies at an accuracy of 0.5 \wn.
This sets the standard uncertainty for F-levels also in HD and D$_2$ isotopologues.

Similar results for \F\ level energies in HD are presented in Table~\ref{tab:Fstate-HD-F0} for the F0 manifold up to $J=26$, in Table~\ref{tab:Fstate-HD-F1} for the F1 manifold up to $J=22$, while in Table~\ref{tab:Fstate-HD-others} some additional levels pertaining to F2 and F3 are listed.

Results for experimental \F\ level energies in D$_2$ are presented in Table~\ref{tab:Fstate-D2-F0} for the F0 manifold observed up to $J=29$, in Table~\ref{tab:Fstate-D2-F1} for the F1 manifold also observed up to $J=29$, while in Table~\ref{tab:Fstate-D2-other} some levels pertaining to the F2, F3 and F4 manifolds are listed.

\begin{table}
\caption{\label{tab:Fstate-HD-F1}
Derived experimental level energies of F1 states in HD measured in this work. A comparison is made with the present non-adiabatic calculations.
}
\begin{tabular}{cc..}
State & $J$& \multicolumn{1}{c}{Exp.}& \multicolumn{1}{c}{NA. Calc.}      \\
\hline
\multirow{17}{*}{F1} &  2  & 100645.7(5) & 100645.39 \\
&  3  & 100673.1(5) & 100672.47 \\
&  4\footnote{Tentative assignment}  & 100706.0(5) & 100708.55 \\
&  5  & 100753.9(5) & 100753.57 \\
&  6  & 100807.5(5) & 100807.50 \\
&  7  & 100870.9(5) & 100870.28 \\
&  8  & 100942.6(5) & 100941.86 \\
&  9  & 101022.1(5) & 101022.16 \\
&  12 & 101314.9(5) & 101314.57 \\
&  13 & 101429.1(5) & 101428.90 \\
&  14 & 101552.9(5) & 101551.50 \\
&  15 & 101681.6(5) & 101682.25 \\
&  16 & 101821.1(5) & 101821.03 \\
&  17 & 101967.4(5) & 101967.72 \\
&  18 & 102123.3(5) & 102122.19 \\
&  19 & 102283.9(5) & 102284.31 \\
&  22 & 102816.3(5) & 102815.11 \\
\hline
\end{tabular}
\end{table}

\begin{table}
\caption{\label{tab:Fstate-HD-others}
Derived experimental level energies of F2 and F3 states in HD measured in this work. A comparison is made with the present non-adiabatic calculations.
}
\begin{tabular}{cc..}
State & $J$& \multicolumn{1}{c}{Exp.}& \multicolumn{1}{c}{NA. Calc.}      \\
\hline
\multirow{7}{*}{F2} & 4  & 101707.9(5) & 101708.55 \\
 & 5  & 101753.0(5) & 101753.57 \\
 & 7  & 101868.3(5) & 101867.56 \\
 & 8  & 101937.5(5) & 101937.25 \\
 & 10 & 102102.9(5) & 102103.56 \\
 & 14 & 102534.9(5) & 102535.81 \\
 & 16 & 102799.9(5) & 102800.02 \\
\hline
\multirow{2}{*}{F3} & 0  & 102579.7(5) & 102577.10 \\
 & 1  & 102586.7(5) & 102585.94 \\
\hline
\end{tabular}
\end{table}

\begin{table}
\caption{\label{tab:Fstate-D2-F0}
Derived experimental level energies of F0 states in \Dm\ measured in this work. A comparison is made with the present non-adiabatic calculations and with the results from FT-emission studies~\cite{Salumbides2014b}.
}
\begin{tabular}{cc...}
State & $J$& \multicolumn{1}{c}{Exp.}& \multicolumn{1}{c}{NA. Calc.} & \multicolumn{1}{c}{Ref.~\cite{Salumbides2014b}.}     \\
\hline
\multirow{26}{*}{F0}&  1 & 99835.2(5)  & 99834.96  & 99835.190(5) \\
&  2 & 99846.6(5)  & 99847.26  & 99847.581(5) \\
&  4 & 99890.1(5)  & 99890.27  & 99890.577(5) \\
&  5 & 99921.4(5)  & 99920.96  & 99921.241(5) \\
&  6 & 99958.1(5)  & 99957.73  &              \\
&  7 & 99999.9(5)  & 100000.60 &              \\
&  8 & 100049.7(5) & 100049.50 &              \\
&  9 & 100104.1(5) & 100104.42 &              \\
&  10 & 100165.0(5) & 100165.33 &              \\
&  11 & 100231.9(5) & 100232.19 &              \\
&  12 & 100305.4(5) & 100304.96 &              \\
&  13 & 100384.0(5) & 100383.60 &              \\
&  14 & 100466.5(5) & 100468.06 &              \\
&  15 & 100558.9(5) & 100558.29 &              \\
&  16 & 100653.5(5) & 100654.24 &              \\
&  17 & 100755.7(5) & 100755.86 &              \\
&  18 & 100862.9(5) & 100863.08 &              \\
&  19 & 100975.4(5) & 100975.84 &              \\
&  20 & 101093.7(5) & 101094.10 &              \\
&  21 & 101217.9(5) & 101217.77 &              \\
&  22 & 101347.6(5) & 101346.78 &              \\
&  23 & 101481.2(5) & 101481.07 &              \\
&  24\footnote{Tentative assignment} & 101618.7(5) & 101620.57 &              \\
&  26 & 101915.7(5) & 101914.87 &              \\
&  27 & 102068.8(5) & 102069.51 &              \\
&  29 & 102392.3(5) & 102393.38 &              \\
\hline
\end{tabular}
\end{table}

\begin{table}
\caption{\label{tab:Fstate-D2-F1}
Derived experimental level energies of F1 states in \Dm\ measured in this work. A comparison is made with the present non-adiabatic calculations and with the results from FT-emission studies~\cite{Salumbides2014b}.
}
\begin{tabular}{cc...}
State & $J$& \multicolumn{1}{c}{Exp.}& \multicolumn{1}{c}{NA. Calc.} & \multicolumn{1}{c}{Ref.~\cite{Salumbides2014b}.}     \\
\hline
\multirow{24}{*}{F1}& 0 & 100687.3(5) & 100685.92 & 100686.239(5)  \\
& 2 & 100703.5(5) & 100704.10 & 100704.409(5)  \\
& 3 & 100721.1(5) & 100722.28 & 100722.567(5)  \\
& 4 & 100746.9(5) & 100746.46 & 100746.741(5)  \\
& 5 & 100777.1(5) & 100776.68 & 100776.969(5)  \\
& 6 & 100812.3(5) & 100812.90 & 100814.897(10) \\
& 7 & 100856.0(5) & 100855.10 &                \\
& 8 & 100903.4(5) & 100903.24 &                \\
& 10 & 101016.7(5) & 101017.27 &                \\
& 11 & 101082.7(5) & 101083.07 &                \\
& 12 & 101155.0(5) & 101154.68 &                \\
& 13 & 101231.8(5) & 101232.05 &                \\
& 14 & 101313.4(5) & 101315.13 &                \\
& 15 & 101403.9(5) & 101403.87 &                \\
& 16 & 101498.8(5) & 101498.23 &                \\
& 17 & 101598.6(5) & 101598.13 &                \\
& 19 & 101813.0(5) & 101814.36 &                \\
& 20 & 101930.1(5) & 101930.55 &                \\
& 21 & 102051.9(5) & 102052.03 &                \\
& 22 & 102179.0(5) & 102178.74 &                \\
& 23 & 102310.0(5) & 102310.61 &                \\
& 24 & 102447.2(5) & 102447.55 &                \\
& 26 & 102735.5(5) & 102736.36 &                \\
& 29 & 103204.8(5) & 103205.68 &                \\
\hline
\end{tabular}
\end{table}

\begin{table}
\caption{\label{tab:Fstate-D2-other}
Derived experimental level energies of F2, F3 and F4 states in \Dm\ measured in this work. A comparison is made with the present non-adiabatic calculations and with the results from FT-emission studies~\cite{Salumbides2014b}.
}
\begin{tabular}{cc...}
State & $J$& \multicolumn{1}{c}{Exp.}& \multicolumn{1}{c}{NA. Calc.} & \multicolumn{1}{c}{Ref.~\cite{Salumbides2014b}.}     \\
\hline
\multirow{8}{*}{F2} & 0  & 101515.1(5) & 101515.76 & 101516.074(5) \\
& 2  & 101534.3(5) & 101533.69 & 101533.990(5) \\
& 3  & 101549.1(5) & 101551.60 & 101551.890(2) \\
& 4  & 101575.4(5) & 101575.46 & 101575.746(2) \\
& 5  & 101604.1(5) & 101605.60 & 101605.675(2) \\
& 11 & 101907.9(5) & 101907.22 &               \\
& 12 & 101978.1(5) & 101977.75 &               \\
& 15 & 102225.6(5) & 102223.10 &               \\
\hline
F3 & 8  & 102529.5(5) & 102529.72 &               \\
\hline
F4 & 7  & 103254.7(5) & 103254.90 &               \\
\hline
\end{tabular}
\end{table}

\section{Non-adiabatic calculations}
\label{NA-calc}

To assist the assignment of the 2+1 REMPI spectra for the F-X system, theoretical evaluations for the level energies of \F\ ro-vibrational levels are used, besides those of the highly accurate levels of the \X\ ground state available from non-adiabatic perturbation theory (NAPT) from the H2SPECTRE program suite~\cite{Czachorowski2018,SPECTRE2022}.
The latter are accurate to $10^{-3}$ \wn, and can be considered exact for the purpose of comparison with the present experimental data.

The accuracy of ab initio computation of excited state level energies has not reached the same level of accuracy as those for the ground state.
However, an extremely accurate calculation was performed in a pre-Born-Oppenheimer framework including relativistic and high-order QED-corrections finding excellent agreement within 0.001 \wn\ for the $J \leq 5$ levels in the \E\ inner well state~\cite{Ferenc2019}.
The Dressler group had previously computed accurate values for the level energies in a non-adiabatic framework~\cite{Quadrelli1990,Yu1994}.
These calculations had reached sub-\wn\ accuracy, and included levels in the \F\ outer well, but the results were limited to low-$J$ values.

\begin{figure}
\begin{center}
\includegraphics[width=\linewidth]{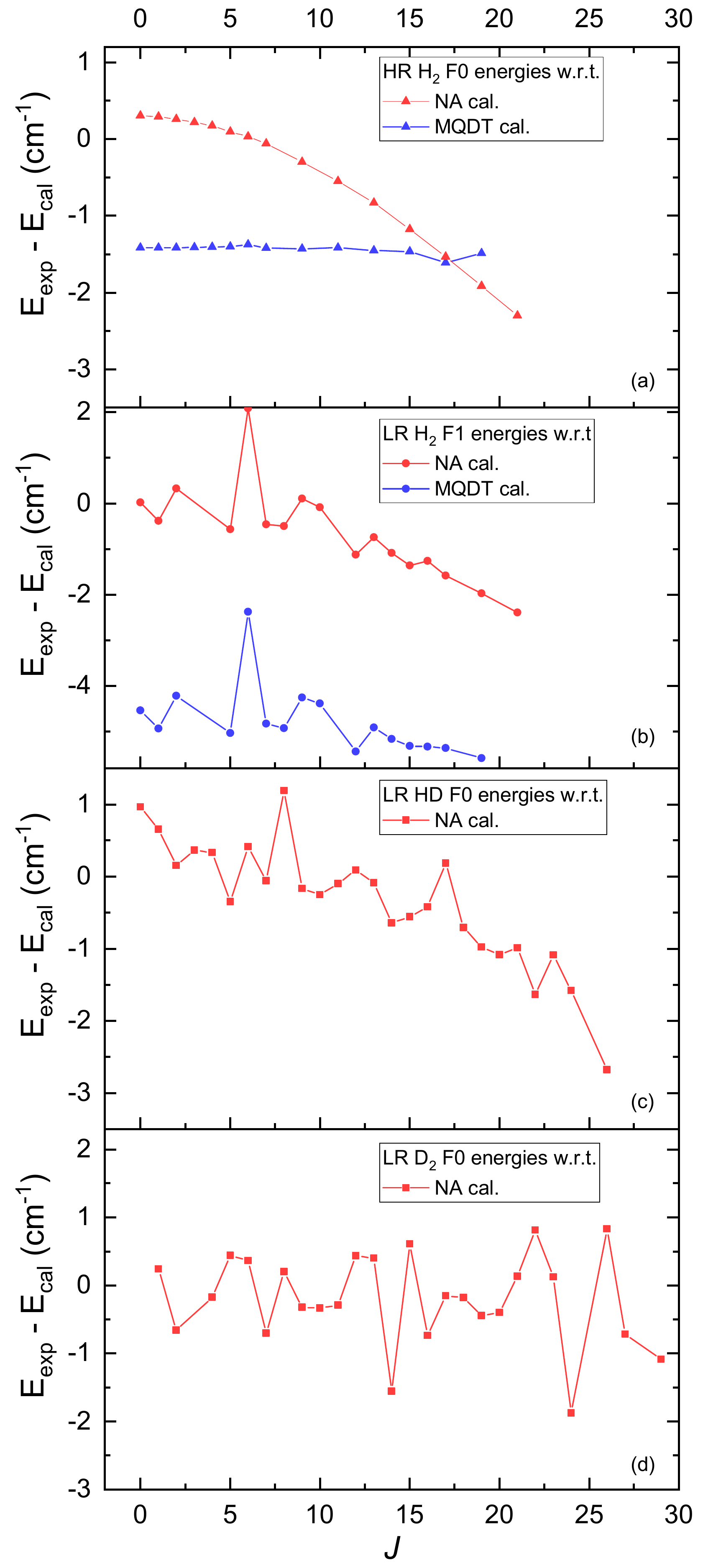}
\caption{\label{H2_EF_diff}
Comparison of the experimentally determined level energies in the various hydrogen isotopologues with calculated values.
(a) Experimental values for the F0 levels in H$_2$ as obtained via the high-resolution (HR) setup with the three-laser scheme, compared with values from the present nonadiabatic calculations and with values from MQDT calculations~\cite{Dickenson2012a}; (b) Experimental values for the F1 levels in H$_2$ obtained via the low-resolution setup with the two-color laser scheme, compared with nonadiabatic and MQDT calculations; Experimental F0 levels in HD (c) and D$_2$ (d) compared with nonadiabatic calculations.}
\end{center}
\end{figure}

In an alternative scheme, following the methods of multi-channel quantum-defect theory (MQDT) an accuracy at the level of few \wn\ was achieved for reported levels in the \E\ inner well of H$_2$~\cite{Dickenson2012a}.
Comparison with the present accurate determination of energies for F-levels shows that for F0-levels up to $J=13$ the MQDT-calculations yield values that are all higher by $1.4$ \wn, and smoothly increasing to an excess of 1.6 \wn\ at $J=19$.
For the low-$J$ levels of F1 the MQDT-values are consistently higher by 4.3 \wn. These comparisons with MQDT results for H$_2$ are displayed in Fig.~\ref{H2_EF_diff}.

For the present assignments and analysis ab initio calculations are performed for ro-vibrational level energies the \EF\ double-well state. We employ the coupled-channel approach described in Ref.~\cite{Quadrelli1990} which includes the first five excited $^1\Sigma_g^+$ states. The term values of the EF($v<6,J=0-5$) states could be reproduced using the original ab initio data obtained by Wolniewicz and coworkers for the BO potential energy curves, adiabatic corrections and homogeneous electronic coupling functions~\cite{Yu1994}.
For the term values presented in this work, the more accurate BO potential energy curves calculated by Silkowski and coworkers \cite{Silkowski2021} were used, which lowered the energies by about 1~cm$^{-1}$.
Heterogeneous couplings between the EF state and excited $^1\Pi_g^+$ were neglected, due to a lack of ab inito data for $\left< ^1\Pi_g^+ | L^+ | {\rm EF} \right>$.
This seems justified in view of the 2s$\sigma$ character of the EF state and taking into account that a pure precession model predicts only mixing between 3d$\sigma$-3d$\pi$.
For the outer-well F levels with low vibrational excitation, no appreciable 3d character is expected and test calculations including the GK-I heterogeneous coupling result in energy shifts below 0.1 \wn\ even for $J=23$ in H$_2$.
Relativistic and QED corrections for E($v=0$) were found to be of the order of 0.1 \wn~\cite{Ferenc2019} and were therefore neglected in the current work.

Results of these calculations are included in the various Tables with experimental data on F-levels for H$_2$ (Tables~\ref{tab:Fstate-H2} and \ref{tab:Fstate-H2-lowres}), for HD (Tables~\ref{tab:Fstate-HD-F0}, \ref{tab:Fstate-HD-F1}, and \ref{tab:Fstate-HD-others}) and D$_2$ (Tables~\ref{tab:Fstate-D2-F0}, \ref{tab:Fstate-D2-F1}, and \ref{tab:Fstate-D2-other}).

The calculations of the level energies of the \F\ outer well states can be compared to the level energies assigned  from experiment.
For this purpose in Fig.~\ref{H2_EF_diff} a comparison is made between the experimental and theoretical values as obtained for two data sets pertaining to H$_2$.
The first panel shows for the high-resolution data in the rotational manifold for F0, a constant offset for the MQDT results, and a gradually increasing offset for the present nonadiabatic calculations as a function of $J$.
The smoothness of the curves indicates that there are no outliers that should be interpreted as misassignments.

The second panel of Fig.~\ref{H2_EF_diff} shows the data for the F1 manifold as obtained from the low-resolution spectra.
An outlier is found for the nonadiabatic calculations of the F1($J=6$) level, which is off by some 2 \wn, while all other deviations smoothly follow a decreasing trend from low $J$ up to a deviation of -3.6 \wn\ at $J=21$. A similar offset is found in the comparison with MQDT data.
In view of this rather strong deviation the assignment of the F1($J=6$) level in Table~\ref{tab:Fstate-H2-lowres} is marked as tentative. Indeed, this level was observed in only one transition.

A similar deviation is found for the F1($J=4$) level in HD, displayed in Fig.~\ref{H2_EF_diff}(c), where the calculated value is higher by 2.8 \wn, while all other deviations are below 1 \wn.
This assignment for the F0($J=24$) level in D$_2$, see  Fig.~\ref{H2_EF_diff}(d), is also marked as tentative.

The $J$-dependent deviation between experimental and non-adiabatic term values, especially for H$_2$ and HD, indicates missing heterogeneous couplings with excited $^1\Pi_g^+$ states, which would lower the theoretical F0 and F1 term values. To improve the coupled-channel calculations, ab inito data for $\left<L^+\right>$ coupling elements is needed.
In contrast, MQDT treats homogeneous and heterogeneous nonadiabatic effects naturally on the same footing and the constant offset in Fig.~\ref{H2_EF_diff} might be reduced drastically by repeating the MQDT calculation with quantum defects extracted from the new ab initio BO potential energy curves.

Recently, the four-body calculations of the inner well E states \cite{Ferenc2019} were extended to the rotational ground state of the outer F well \cite{Saly2023}, resulting in a term value of 99365(1)~cm$^{-1}$ for F(0,0). As can be seen from the first entry in Table~\ref{tab:Fstate-H2}, this deviates by -1.2 \wn\ from the experimental value, compared to a 0.3 \wn\ discrepancy resulting from a comparison with the coupled-channel approach.
While four-body calculations allow ultimately to reach higher accuracies, they remain computationally challenging for large $R$ and high $J$.

\section{Conclusion}

In the present study vibrationally and rotationally excited states in the \X\ electronic ground states of H$_2$, HD and D$_2$ fragment species were produced via two-photon ultraviolet photolysis of H$_2$S, HDS and D$_2$S, respectively.
These states, with wave function density at large internuclear separation, were subjected to laser spectroscopic investigations of the \F\ - \X\ system in the hydrogen isotopologues.
Combined with known accurate theoretical values for the X($v,J$) levels~\cite{Czachorowski2018,Pachucki2022,SPECTRE2022}, level energies of series of rotational states in the \F\ outer well state were determined up to high angular momentum.
In H$_2$ results were obtained via a high-resolution three-color laser scheme for $v_F=0, 1$ at typical uncertainties of $5 \times 10^{-3}$ \wn.
In addition, from a lower resolution two-color laser scheme additional $v_F=0, 1$ levels as well as $v_F=2$ levels were determined at an accuracy of 0.5 \wn.
For HD and D$_2$ energies of many F-levels $v_F=0-4$ were determined for the first time.

Full non-adiabatic term values for the \F\ outer well levels were calculated, based on a coupled-channel approach including $^1\Sigma_g^+$ states.
Good agreement, at the level of the experimental accuracy of 0.5 \wn,  is obtained from these calculations.
%that prove to be more accurate than previously obtained theoretical results from MQDT calculations.

\clearpage

%\bibliography{Sulphur,Lasers+Spectroscopy}
%\bibliography{Hydrogen,Lasers+Spectroscopy,S,QB,H2ion}
%

%\newpage
%\appendix

\section{Appendix}

In this Appendix line lists will be provided on the measured frequencies of the \F\ - \X\ two-photon transitions in all three isotopologues, Table~\ref{H2list} for H$_2$, Table~\ref{HDlist} for HD, and Table~\ref{D2list} for D$_2$.
The listed data are acquired via the low-resolution two-color scheme for which an uncertainty of 0.5 \wn\ is determined.

\begin{longtable}{cccccccc}
\caption{\Hm\ linelist}
\label{H2list}\\
\hline
$v_\text{F}$ & $v_\text{X}$ & Line & Freq. & $v_\text{F}$ & $v_\text{X}$ & Line & Freq. \\
\hline
\endfirsthead
\caption{\Hm\ linelist (cont.)}\\
\hline
$v_\text{F}$ & $v_\text{X}$ & Line & Freq. & $v_\text{F}$ & $v_\text{X}$ & Line & Freq.\\
\hline
\endhead
\hline
(cont.)
\endfoot
\hline
\endlastfoot
0  & 6  & O(23)  & 65178.27 &  0  & 6  & Q(23)  & 65678.80   \\
0  & 7  & Q(19)  & 66470.39 &  0  & 7  & Q(20)  & 65933.05  \\
0  & 7  & O(21)  & 64978.92 &  0  & 7  & Q(21)  & 65441.88  \\
0  & 7  & S(21)  & 65939.95 &  0  & 8  & O(17)  & 65662.00  \\
0  & 8  & Q(17)  & 66044.80 &  0  & 8  & Q(18)  & 65579.83  \\
0  & 8  & O(19)  & 64734.17 &  0  & 8  & Q(19)  & 65157.61  \\
0  & 8  & S(19)  & 65619.98 &  0  & 9  & Q(13)  & 66409.96  \\
0  & 9  & Q(14)  & 65983.54 &  0  & 9  & O(15)  & 65231.64   \\
0  & 9  & Q(15)  & 65571.25 &  0  & 9  & S(15)  & 65954.46   \\
0  & 9  & Q(16)  & 65185.48 &  0  & 9  & S(16)  & 65587.59  \\
0  & 9  & O(17)  & 64456.20 &  0  & 9  & Q(17)  & 64836.51   \\
0  & 9  & S(17)  & 65260.54 &  0  & 10 & Q(9)   & 66400.53   \\
0  & 10 & Q(10)  & 66067.28 &  0  & 10 & O(11)  & 65476.88  \\
0  & 10 & Q(11)  & 65728.25 &  0  & 10 & S(11)  & 66024.40   \\
0  & 10 & O(12)  & 65116.96 &  0  & 10 & Q(12)  & 65392.15   \\
0  & 10 & O(13)  & 64768.88 &  0  & 10 & Q(13)  & 65065.84   \\
0  & 10 & S(13)  & 65406.04 &  0  & 10 & O(14)  & 64443.43  \\
0  & 10 & Q(14)  & 64761.93 &  0  & 10 & S(14)  & 65123.36  \\
0  & 10 & O(15)  & 64152.75 &  0  & 10 & Q(15)  & 64493.97  \\
0  & 10 & S(15)  & 64872.86 &  0  & 11 & Q(5)   & 65931.85   \\
0  & 11 & Q(7)   & 65510.35 &  0  & 11 & O(8)   & 65093.42  \\
0  & 11 & Q(8)   & 65274.82 &  0  & 11 & S(9)   & 65281.71   \\
0  & 11 & Q(10)  & 64786.26 &  0  & 11 & S(10)  & 65059.98   \\
0  & 11 & O(11)  & 64297.61 &  0  & 11 & Q(11)  & 64548.96   \\
0  & 11 & S(11)  & 64845.70 &  0  & 11 & Q(12)  & 64331.06   \\
0  & 12 & Q(3)   & 64889.86 &  0  & 12 & S(3)   & 64998.96   \\
0  & 12 & Q(4)   & 64781.77 &  0  & 12 & Q(5)   & 64654.02   \\
0  & 12 & S(5)   & 64811.24 &  0  & 12 & O(6)   & 64376.07  \\
0  & 12 & Q(6)   & 64511.40 &  0  & 12 & O(7)   & 64199.67   \\
0  & 12 & Q(7)   & 64357.58 &  0  & 12 & S(7)   & 64562.72   \\
0  & 12 & O(8)   & 64024.42 &  0  & 12 & O(9)   & 63849.55   \\
0  & 12 & Q(9)   & 64055.69 &  0  & 12 & S(9)   & 64305.57   \\
0  & 12 & O(10)  & 63696.70 &  0  & 12 & Q(10)  & 63924.40   \\
0  & 13 & Q(3)   & 63905.42 &  0  & 13 & Q(4)   & 63841.87  \\
0  & 13 & O(5)   & 63661.97 &  0  & 13 & Q(5)   & 63771.65   \\
0  & 12 & Q(0)   & 65060.68 &  0  & 12 & Q(1)   & 65032.35  \\
0  & 12 & O(4)   & 64696.16 &  0  & 13 & Q(0)   & 64011.54  \\
0  & 13 & S(0)   & 64051.74 &  0  & 13 & Q(1)   & 63993.60  \\
1  & 6  & O(23)  & 66308.66 &  1  & 7  & O(21)  & 66120.98   \\
1  & 7  & Q(21)  & 66570.35 &  1  & 8  & O(19)  & 65887.53   \\
1  & 8  & Q(19)  & 66299.74 &  1  & 9  & Q(16)  & 66343.24  \\
1  & 9  & Q(17)  & 65990.07 &  1  & 10 & Q(12)  & 66566.23   \\
1  & 10 & Q(13)  & 66236.87 &  1  & 10 & Q(14)  & 65928.48   \\
1  & 10 & Q(15)  & 65655.57 &  1  & 11 & Q(8)   & 66460.64   \\
1  & 11 & Q(9)   & 66214.56 &  1  & 11 & Q(12)  & 65504.24   \\
1  & 12 & Q(9)   & 65238.30 &  1  & 12 & O(10)  & 64881.97   \\
1  & 12 & Q(10)  & 65105.98 &  1  & 13 & S(7)   & 65027.54   \\
1  & 12 & Q(7)   & 65545.67 &  1  & 13 & Q(0)   & 65207.25   \\
1  & 13 & Q(1)   & 65188.76 &  1  & 13 & S(1)   & 65247.84   \\
1  & 13 & Q(2)   & 65151.55 &  1  & 13 & O(3)   & 65038.08   \\
1  & 13 & Q(3)   & 65099.52 &  1  & 13 & O(5)   & 64853.59   \\
1  & 13 & Q(5)   & 64963.13 &  1  & 13 & S(5)   & 65116.96   \\
1  & 13 & Q(6)   & 64892.62 &  1  & 13 & O(7)   & 64670.11   \\
1  & 13 & Q(7)   & 64825.66 &  1  & 14 & O(4)   & 64477.19   \\
2  & 12 & Q(9)   & 66370.16 &  2  & 12 & S(9)   & 66610.39   \\
2  & 13 & S(4)   & 66299.74 &  2  & 13 & O(7)   & 65807.52  \\
2  & 13 & Q(7)   & 65959.96 &  2  & 14 & Q(1)   & 65720.46 \\
2  & 14 & O(2)   & 65675.34 &  2  & 14 & Q(2)   & 65711.49 \\
2  & 14 & S(1)   & 65777.95 &  2  & 14 & Q(3)   & 65700.67  \\
\end{longtable}

%\clearpage

\begin{longtable}{cccccccc}
\caption{HD linelist}
\label{HDlist}\\\\
\hline
$v_\text{F}$ & $v_\text{X}$ & Line & Freq. & $v_\text{F}$ & $v_\text{X}$ & Line & Freq. \\
\hline
\endfirsthead
\caption{HD linelist (cont.)}\\
\hline
$v_\text{F}$ & $v_\text{X}$ & Line & Freq. & $v_\text{F}$ & $v_\text{X}$ & Line & Freq. \\
\hline
\endhead
\hline
(cont.)
\endfoot
\hline
\endlastfoot
0 & 7  & Q(26) & 65910.49 & 0 & 8  & O(24) & 65278.27 \\
0 & 8  & Q(23) & 66131.28 & 0 & 8  & Q(24) & 65677.54 \\
0 & 9  & Q(20) & 66237.79 & 0 & 9  & Q(21) & 65808.52 \\
0 & 9  & Q(22) & 65405.57 & 0 & 9  & Q(23) & 65040.85 \\
0 & 10 & O(18) & 65512.86 & 0 & 10 & Q(17) & 66208.30 \\
0 & 10 & O(19) & 65126.52 & 0 & 10 & Q(18) & 65820.40 \\
0 & 10 & Q(19) & 65449.31 & 0 & 10 & Q(20) & 65101.79 \\
0 & 10 & Q(21) & 64791.43 & 0 & 10 & S(20) & 65472.05 \\
0 & 10 & S(21) & 65177.87 & 0 & 11 & O(14) & 65790.64 \\
0 & 11 & Q(13) & 66369.71 & 0 & 11 & O(15) & 65440.48 \\
0 & 11 & Q(14) & 66032.76 & 0 & 11 & O(16) & 65097.61 \\
0 & 11 & O(17) & 64770.99 & 0 & 11 & S(14) & 66307.97 \\
0 & 11 & Q(16) & 65372.29 & 0 & 11 & O(18) & 64466.84 \\
0 & 11 & Q(17) & 65063.71 & 0 & 11 & S(17) & 65385.50 \\
0 & 11 & Q(19) & 64520.56 & 0 & 11 & S(18) & 65113.32 \\
0 & 12 & O(9)  & 66058.38 & 0 & 12 & Q(9)  & 66212.27 \\
0 & 12 & Q(10) & 65960.18 & 0 & 12 & O(12) & 65224.01 \\
0 & 12 & Q(11) & 65698.30 & 0 & 12 & O(13) & 64942.15 \\
0 & 12 & Q(12) & 65431.69 & 0 & 12 & O(14) & 64664.34 \\
0 & 12 & Q(13) & 65165.40 & 0 & 12 & O(15) & 64400.09 \\
0 & 12 & S(12) & 65673.24 & 0 & 12 & Q(14) & 64905.45 \\
0 & 12 & O(16) & 64157.36 & 0 & 12 & S(13) & 65423.51 \\
0 & 12 & Q(15) & 64658.65 & 0 & 12 & Q(16) & 64433.57 \\
0 & 12 & S(15) & 64951.92 & 0 & 13 & Q(0)  & 66102.10 \\
0 & 13 & Q(2)  & 66019.09 & 0 & 13 & Q(3)  & 65939.12 \\
0 & 13 & Q(4)  & 65833.49 & 0 & 13 & Q(5)  & 65705.22 \\
0 & 13 & Q(6)  & 65557.70 & 0 & 13 & O(8)  & 65074.39 \\
0 & 13 & Q(7)  & 65391.40 & 0 & 13 & O(9)  & 64867.03 \\
0 & 13 & Q(8)  & 65212.15 & 0 & 13 & O(10) & 64655.98  \\
0 & 13 & S(7)  & 65546.36 & 0 & 13 & Q(9)  & 65023.09 \\
0 & 13 & Q(10) & 64825.92 & 0 & 13 & O(12) & 64229.23 \\
0 & 13 & Q(11) & 64629.52 & 0 & 13 & O(13) & 64030.79 \\
0 & 13 & S(10) & 65034.18 & 0 & 13 & Q(12) & 64437.44 \\
0 & 13 & S(11) & 64855.29 & 0 & 13 & Q(13) & 64255.76 \\
0 & 13 & S(12) & 64678.28 & 0 & 13 & Q(14) & 64094.35 \\
0 & 13 & S(13) & 64514.57 & 0 & 14 & O(3)  & 64740.57 \\
0 & 14 & Q(2)  & 64848.77 & 0 & 14 & Q(3)  & 64785.54 \\
0 & 14 & Q(4)  & 64704.94 & 0 & 14 & Q(5)  & 64607.02 \\
0 & 14 & Q(6)  & 64496.31 & 0 & 14 & S(5)  & 64725.84 \\
0 & 14 & Q(7)  & 64374.74 & 0 & 14 & Q(8)  & 64248.71\\
0 & 14 & Q(9)  & 64116.85 & 0 & 14 & Q(10) & 63991.87  \\
0 & 14 & S(9)  & 64307.14 & 0 & 14 & Q(11) & 63876.61 \\
0 & 14 & S(11) & 64101.46 & 0 & 15 & Q(1)  & 63986.80 \\
0 & 15 & Q(2)  & 63957.38 & 0 & 15 & Q(3)  & 63916.32 \\
0 & 15 & Q(4)  & 63863.43 & 0 & 15 & Q(5)  & 63801.30 \\
0 & 15 & Q(6)  & 63736.63 & 1 & 8  & O(24) & 66276.52  \\
1 & 10 & O(20) & 65776.08 & 1 & 10 & Q(19) & 66456.76 \\
1 & 11 & Q(16) & 66390.80 & 1 & 11 & Q(17) & 66076.65 \\
1 & 12 & Q(13) & 66191.40 & 1 & 12 & Q(14) & 65929.67 \\
1 & 12 & Q(17) & 65255.86 & 1 & 12 & S(17) & 65572.00 \\
1 & 13 & Q(8)  & 66248.12 & 1 & 13 & Q(12) & 65465.48 \\
1 & 13 & Q(13) & 65282.80 & 1 & 13 & Q(14) & 65121.73 \\
1 & 13 & S(13) & 65534.97 & 1 & 14 & Q(7)  & 65413.09 \\
1 & 14 & Q(8)  & 65284.36 & 1 & 14 & Q(9)  & 65152.16 \\
1 & 14 & O(11) & 64721.76 & 1 & 14 & Q(12) & 64812.33 \\
1 & 15 & Q(2)  & 64999.29 & 1 & 15 & Q(3)  & 64958.24 \\
1 & 15 & O(5)  & 64761.69 & 1 & 15 & S(2)  & 65059.50 \\
1 & 15 & Q(5)  & 64842.24 & 1 & 15 & Q(6)  & 64774.71 \\
1 & 15 & Q(7)  & 64707.07 & 1 & 15 & Q(8)  & 64643.85 \\
1 & 15 & Q(9)  & 64591.99 & 1 & 16 & Q(3)  & 64413.41 \\
2 & 12 & Q(16) & 66430.36 & 2 & 13 & Q(14) & 66102.40 \\
2 & 14 & O(7)  & 66294.74 & 2 & 14 & Q(7)  & 66410.03 \\
2 & 15 & Q(8)  & 65639.21 & 2 & 15 & S(8)  & 65804.57 \\
2 & 16 & Q(4)  & 65396.35 & 3 & 16 & Q(1)  & 66400.68 \\
3 & 17 & Q(0)  & 66177.53  \\
\end{longtable}

%\clearpage

\begin{longtable}{cccccccc}
\caption{\Dm\ linelist}
\label{D2list}\\
\hline
$v_\text{F}$ & $v_\text{X}$ & Line & Freq. & $v_\text{F}$ & $v_\text{X}$ & Line & Freq. \\
\hline
\endfirsthead
\caption{\Dm\ linelist (cont.)}\\
\hline
$v_\text{F}$ & $v_\text{X}$ & Line & Freq. & $v_\text{F}$ & $v_\text{X}$ & Line & Freq. \\
\hline
\endhead
\hline
(cont.)
\endfoot
\hline
\endlastfoot
0 & 10 & Q(29) & 65682.19 & 0 & 11 & Q(24) & 66509.52 \\
0 & 11 & Q(26) & 65787.37 & 0 & 11 & Q(27) & 65449.94 \\
0 & 12 & Q(20) & 66826.12 & 0 & 12 & Q(21) & 66481.00 \\
0 & 12 & Q(23) & 65812.66 & 0 & 12 & Q(24) & 65494.50 \\
0 & 13 & Q(17) & 66647.29 & 0 & 13 & Q(18) & 66345.49 \\
0 & 13 & Q(19) & 66044.44 & 0 & 13 & Q(20) & 65748.88 \\
0 & 13 & Q(21) & 65460.65 & 0 & 13 & Q(22) & 65184.22 \\
0 & 13 & S(24) & 64978.40 & 0 & 14 & Q(17) & 65588.82 \\
0 & 14 & Q(18) & 65336.29 & 0 & 14 & Q(19) & 65091.05 \\
0 & 14 & S(17) & 65807.90 & 0 & 14 & S(18) & 65567.20 \\
0 & 15 & Q(6)  & 66766.96 & 0 & 15 & Q(7)  & 66627.87 \\
0 & 15 & Q(8)  & 66476.58 & 0 & 15 & Q(9)  & 66308.67 \\
0 & 15 & Q(11) & 65938.80 & 0 & 15 & Q(12) & 65742.01 \\
0 & 15 & Q(13) & 65537.97 & 0 & 15 & O(16) & 64728.82 \\
0 & 15 & Q(14) & 65327.24 & 0 & 15 & Q(15) & 65122.32 \\
0 & 15 & Q(16) & 64914.63 & 0 & 15 & S(14) & 65515.41 \\
0 & 15 & S(17) & 64933.43 & 0 & 15 & Q(20) & 64181.15 \\
0 & 16 & Q(5)  & 65771.20 & 0 & 16 & O(8)  & 65331.49 \\
0 & 16 & Q(6)  & 65668.79 & 0 & 16 & Q(7)  & 65551.35 \\
0 & 16 & Q(8)  & 65421.55 & 0 & 16 & Q(9)  & 65281.49 \\
0 & 16 & S(7)  & 65655.20 & 0 & 16 & O(12) & 64678.32 \\
0 & 16 & Q(10) & 65131.89 & 0 & 16 & Q(11) & 64977.58 \\
0 & 16 & Q(12) & 64816.94 & 0 & 16 & S(10) & 65272.84 \\
0 & 16 & Q(13) & 64655.44 & 0 & 16 & S(18) & 64166.23 \\
0 & 16 & Q(4)  & 65859.20 & 0 & 17 & Q(5)  & 64825.02 \\
0 & 17 & Q(6)  & 64740.84 & 0 & 17 & S(7)  & 64749.95 \\
0 & 17 & O(12) & 63947.24 & 0 & 17 & Q(11) & 64203.54 \\
0 & 17 & Q(12) & 64088.77 & 0 & 17 & Q(2)  & 65000.13 \\
0 & 17 & Q(4)  & 64897.35 & 0 & 18 & S(4)  & 64185.23 \\
0 & 18 & Q(1)  & 64219.44 & 0 & 18 & Q(2)  & 64195.02 \\
0 & 18 & S(2)  & 64238.44 & 0 & 20 & S(3)  & 63267.84 \\
1 & 10 & Q(29) & 66494.72 & 1 & 11 & Q(26) & 66608.29 \\
1 & 12 & Q(23) & 66641.46 & 1 & 12 & Q(24) & 66322.34 \\
1 & 13 & Q(20) & 66584.83 & 1 & 13 & Q(21) & 66294.44 \\
1 & 13 & Q(22) & 66015.62 & 1 & 14 & Q(16) & 66688.32 \\
1 & 14 & Q(17) & 66431.15 & 1 & 14 & Q(19) & 65928.48 \\
1 & 14 & Q(20) & 65691.85 & 1 & 15 & Q(11) & 66789.99 \\
1 & 15 & Q(12) & 66591.64 & 1 & 15 & Q(13) & 66385.98 \\
1 & 15 & Q(15) & 65967.18 & 1 & 15 & Q(19) & 65179.15 \\
1 & 16 & Q(11) & 65828.38 & 1 & 16 & Q(14) & 65339.16 \\
1 & 16 & Q(17) & 64895.09 & 1 & 16 & Q(4)  & 66717.04 \\
1 & 16 & Q(6)  & 66523.54 & 1 & 16 & Q(7)  & 66406.87 \\
1 & 16 & S(6)  & 66614.36 & 1 & 17 & O(12) & 64800.23 \\
1 & 17 & Q(10) & 65171.10 & 1 & 17 & Q(11) & 65052.64 \\
1 & 17 & Q(12) & 64937.89 & 1 & 17 & Q(15) & 64624.84 \\
1 & 18 & S(8)  & 64831.26 & 1 & 18 & S(9)  & 64769.49 \\
1 & 18 & Q(2)  & 65052.23 & 1 & 18 & Q(3)  & 65017.30 \\
1 & 18 & Q(4)  & 64973.60 & 1 & 18 & Q(5)  & 64920.35 \\
1 & 18 & Q(6)  & 64858.44 & 1 & 18 & Q(7)  & 64790.97 \\
1 & 18 & Q(8)  & 64717.59 & 1 & 19 & O(8)  & 64170.87 \\
1 & 20 & Q(0)  & 64082.43 & 2 & 17 & Q(5)  & 66507.91 \\
2 & 17 & Q(11) & 65879.61 & 2 & 17 & Q(12) & 65760.99 \\
2 & 17 & Q(15) & 65446.28 & 2 & 18 & Q(2)  & 65883.04 \\
2 & 18 & Q(3)  & 65845.22 & 2 & 18 & Q(4)  & 65802.74 \\
2 & 19 & Q(4)  & 65229.39 & 2 & 20 & Q(0)  & 64910.20 \\
3 & 19 & Q(8)  & 65889.37 & 4 & 19 & S(5)  & 66846.32 \\
\end{longtable}

\end{document}